%% file: main.tex
\documentclass[runningheads]{llncs}
\usepackage[T1]{fontenc}
\usepackage[utf8]{inputenc}
\usepackage{environ}
\usepackage{amsfonts}
\usepackage{amssymb}
\usepackage{amsmath}
\usepackage{mathtools}
\usepackage{graphicx}
\usepackage{stackengine}
\usepackage{tabularx}
\usepackage{booktabs}
\usepackage{xspace}
\usepackage{multirow}
\usepackage{cases}
\usepackage{subcaption}
\usepackage[inline]{enumitem}
\usepackage{xcolor}
\usepackage{microtype}%if unwanted, comment out or use option "draft"

\usepackage{tikz}
\usetikzlibrary{calc,positioning,decorations.pathreplacing,fit}
\usepackage{circuitikz}
\usepackage{xcolor}
\usepackage{hyperref}
\hypersetup{
  colorlinks = true,
  urlcolor = blue,
  linkcolor = blue,
  citecolor =red
}

\usepackage[shortcuts]{extdash}

\input macros

\usepackage[sort&compress,numbers]{natbib}

\newcommand{\ignore}[1]{}

\title{(Near)-Optimal Algorithms for Sparse Separable Convex Integer Programs}
\titlerunning{(Near)-Optimal Algorithms for Sparse Separable Convex IPs}
\author{
  Christoph Hunkenschröder\inst{1}\orcidID{0000-0001-5580-3677}
  \and
  Martin Kouteck\'y\inst{3}\orcidID{0000-0002-7846-0053}
  \and
  Asaf Levin\inst{2}\orcidID{0000-0001-7935-6218}
  \and
  Tung Anh Vu\inst{3}\orcidID{0000-0002-8902-5196}
}
\authorrunning{C. Hunkenschröder et al.}
\institute{%
  TU Berlin, Berlin, Germany \\
  \email{hunkenschroeder@tu-berlin.de}
  \and
  Technion -- Israel Institute of Technology, Haifa, Israel \\
  \email{levinas@ie.technion.ac.il}
  \and
  Charles University, Prague, Czech Republic \\
  \email{\{koutecky,tung\}@iuuk.mff.cuni.cz}
}
\begin{document}

\maketitle

\input abstract

\input intro

% \input overview

\input preliminaries
\input scaling-algorithm

\input primal-algorithm

\input sensitivity

\input dual-algorithm

\vspace{-1em}

\input conclusion

\begin{credits}
  \subsubsection{\ackname}
  Hunkenschröder acknowledges support by the Einstein Foundation Berlin.
  Levin is partially supported by ISF - Israel Science Foundation grant number 1467/22.
  Koutecký and Vu are partially supported by Charles Univ. project UNCE 24/SCI/008, by the ERC-CZ project LL2406 of the Ministry of Education of Czech Republic.
  Koutecký is additionally supported by projects 22-22997S and 25-17221S of GA ČR, Vu by the project 24-10306S of GA ČR.
\end{credits}

\clearpage

\begin{appendix}

\appendixText

% \input n-fold-hochbaum

\input convolution-tree-opt

\end{appendix}

\bibliographystyle{splncs04}
\bibliography{psp}
%\bibliography{soda24-refs}

\end{document}

%% file: macros.tex
\DeclarePairedDelimiter\ceil{\lceil}{\rceil}
\DeclarePairedDelimiter\floor{\lfloor}{\rfloor}

\newcommand\II{\mathcal{I}}
\newcommand\JJ{\mathcal{J}}
\newcommand{\df}{\mathrel{\mathop:}=}

\newcommand{\hy}{\hbox{-}\nobreak\hskip0pt}
\def\N{\mathbb{N}}
\def\R{\mathbb{R}}
\def\Z{\mathbb{Z}}

\def \G {\mathcal{G}}
\def \T {\mathcal{T}}

\DeclareMathOperator{\suppo}{\mathrm{supp}}
\DeclareMathOperator{\Ker}{\mathrm{Ker}}

\def \Oh {\mathcal{O}}

\def \PP {\mathcal{P}}

\def \t {^{^\intercal}}

\def\ve#1{%
  \mathchoice{\mbox{\boldmath$\displaystyle\bf#1$}}
  {\mbox{\boldmath$\textstyle\bf#1$}}
  {\mbox{\boldmath$\scriptstyle\bf#1$}}
  {\mbox{\boldmath$\scriptscriptstyle\bf#1$}}
}
\newcommand\vea{{\ve a}}
\newcommand\vealpha{{\ve \alpha}}
\newcommand\veb{{\ve b}}

\newcommand\vece{{\ve e}}

\newcommand\veg{{\ve g}}
\newcommand\veG{{\ve G}}
\newcommand\veh{{\ve h}}

\newcommand\vel{{\ve l}}

\newcommand\ver{{\ve r}}
\newcommand\ves{{\ve s}}

\newcommand\veu{{\ve u}}
\newcommand\vev{{\ve v}}
\newcommand\vew{{\ve w}}
\newcommand\vex{{\ve x}}
\newcommand\vey{{\ve y}}
\newcommand\vez{{\ve z}}
\newcommand{\norm}[1]{\left\| #1 \right\|}

\newcommand\vegamma{{\ve \gamma}}
\newcommand\vedelta{{\ve \delta}}
\newcommand\vebeta{{\ve \beta}}

\newcommand\vezero{{\ve 0}}

\newcommand{\NPh}{$\mathsf{NP}$\hy\textsf{hard}\xspace}

\newcommand{\FPT}{$\mathsf{FPT}$\xspace}

\newcommand{\bigO}{\mathcal{O}}
\newcommand{\PN}[1]{\textsc{#1}} % Problem Name

\DeclareMathOperator{\convol}{\mathrm{convol}}

\def \sign {{\rm sign}}
\def \undffd {\bot}

\DeclareMathOperator{\td}{\mathrm{td}}
\DeclareMathOperator{\ttd}{\mathrm{th}}

\DeclareMathOperator{\cl}{\mathrm{cl}}
\DeclareMathOperator{\height}{\mathrm{height}}
\DeclareMathOperator{\gap}{\mathrm{gap}}
\DeclareMathOperator{\Sol}{\mathrm{Sol}}

\usepackage{etoolbox}

\newcommand{\sv}[1]{}

\newcommand{\appendixText}{}

\newcommand{\toappendix}[1]{#1}

\newcommand{\replabel}{\label} % will be redefined in restatements

\ExplSyntaxOn

\NewDocumentCommand{\repeattheorem}{m}
 {
  \group_begin:
  \renewcommand{\thetheorem}{\ref{#1}}
  \renewcommand{\replabel}[1]{\tag{\ref{##1}}}
  \prop_item:Nn \g_reptheorem_prop { #1 }
  \endtheorem
  \group_end:
 }

\NewDocumentEnvironment{reptheorem}{m+b}
 {
  \prop_gput:Nnn \g_reptheorem_prop { #1 } { \theorem #2 \endtheorem }
  \theorem#2\unskip\label{#1}\endtheorem
 }{}

\prop_new:N \g_reptheorem_prop

\ExplSyntaxOff

%% file: abstract.tex
\begin{abstract}
  We study the general integer programming (IP) problem of optimizing a separable convex function over the integer points of a polytope: $\min \{f(\vex) \mid A\vex = \veb, \, \vel \leq \vex \leq \veu, \, \vex \in \Z^n\}$.
  The number of variables $n$ is a variable part of the input, and we consider the regime where the constraint matrix $A$ has small coefficients $\|A\|_\infty$ and small primal or dual treedepth $\td_P(A)$ or $\td_D(A)$, respectively.
  Equivalently, we consider block-structured matrices, in particular $n$-fold, tree-fold, $2$-stage and multi-stage matrices.

  We ask about the possibility of near-linear time algorithms in the general case of (non-linear) separable convex functions.
  The techniques of previous works for the linear case are inherently limited to it;
  in fact, no strongly-polynomial algorithm may exist due to a simple unconditional information-theoretic lower bound of $n \log \|\veu-\vel\|_\infty$, where $\vel, \veu$ are the vectors of lower and upper bounds.
  Our first result is that with parameters $\td_P(A)$ and $\|A\|_\infty$, this lower bound can be matched (up to dependency on the parameters). Second, with parameters $\td_D(A)$ and $\|A\|_\infty$, the situation is more involved, and we design an algorithm with time complexity $g(\td_D(A), \|A\|_\infty) n \log(n) \log \|\veu-\vel\|_\infty$ where~\(g\) is some computable function.
  We conjecture that a stronger lower bound is possible in this regime, and our algorithm is in fact optimal.

  %  In designing new algorithms, we focus on how their time complexity depends on the dimension $n$ and the lower and upper bounds $\vel, \veu$.
  %  There is an unconditional lower bound of $n \log \|\veu-\vel\|_\infty$ and our goal is to match it.
  %  Our new algorithms have time complexities which depend on $n$ and $\vel, \veu$ as either $n \log \|\veu-\vel\|_\infty$ or $n \log n \log \|\veu-\vel\|_\infty$ for the case of primal and dual treedepth, respectively.
  Our algorithms combine ideas from scaling, proximity, and sensitivity of integer programs, together with a new dynamic data structure.

\keywords{integer programming, parameterized complexity, Graver basis, treedepth, \(n\)-fold, tree-fold, 2-stage stochastic, multi-stage stochastic}% mandatory: Please provide 1-5 keywords
\end{abstract}

%% file: intro.tex
% TEX root = ./main.tex
\section{Introduction}
Our focus is on the integer (linear) programming problem in standard form
\begin{gather}
  \min\left\{f(\vex) \mid A\vex=\veb,\, \vel\leq\vex\leq\veu,\, \vex\in\Z^{n}\right\} \tag{IP}, \text{ and} \label{IP}
  \\
  \min\left\{\vew \vex \mid A\vex=\veb,\, \vel\leq\vex\leq\veu,\, \vex\in\Z^{n}\right\} \tag{ILP}, \label{ILP}
\end{gather}
with $A \in \Z^{m \times n}$, $\veb\in\Z^m$, $\vel,\veu, \vew\in \Z^n$, and $f: \R^n \to \R$ a separable convex function, that is, $f$ can be expressed as $f(\vex) = \sum_{i=1}^n f_i(x_i)$ and each $f_i: \R \to \R$ is convex.
We write vectors in boldface (e.g., $\vex, \vey$) and their entries in normal font (e.g., the $i$\hy{}th entry of~$\vex$ is~$x_i$).
Throughout this paper we shall follow the notation of Grötschel, Lovász, and Schrijver~\cite{GLS} and write the dot product of vectors~\(\vew\) and~\(\vex\) as~\(\vew\vex\) instead of~\(\vew\t\vex\).
In this work the objective function~\(f\) is represented by a comparison oracle: given two points $\vex, \vey \in \R^n$, it returns the answer to the query~$f(\vex ) > f(\vey)$ in constant time.
%\tung{I changed it to comparison oracle and defined it. OK?}
% We compare vectors entry-wise, i.e.~if~\(\vex\) and~\(\vey\) are two vectors of length~\(n\), then~\(\vex \le \vey\) if for every~\(i \in [n]\) we have~\(x_i \le y_i\).
%Our results can be extended to the setting where some lower and upper are infinite by methods described in~\cite[Section 3.5]{framework}), but for simplicity, we focus on finite bounds here.\martin{move to a later point in the paper}
\ref{IP}~is a fundamental optimization problem with rich theory and many practical applications~\cite{BrielVossenKambhampati05,VossenBallLotemNau99,Toth01,FloudasLin05,LodiMM02,AlumurK08,DBLP:journals/informs/AchterbergBGRW20,DBLP:journals/informs/KnuevenOW20,DBLP:journals/mp/AndersonHMTV20}.
%\byasaf{Add a reference from the last 5 years}
%\byasaf{Say something about encoding of~\(f\)}

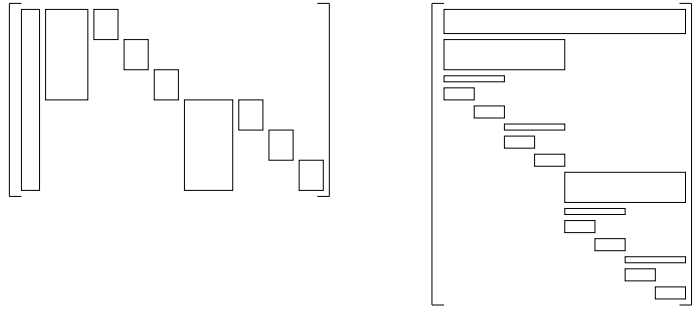
\begin{figure}[ht]
  \centering
  \resizebox{0.75\textwidth}{!}{
    \begin{circuitikz}
      \tikzstyle{every node}=[font=\LARGE]
      \draw [, line width=0.5pt ] (2.5,12.25) rectangle (3.25,4.75);
      \draw [, line width=0.5pt ] (3.5,12.25) rectangle (5.25,8.5);
      \draw [, line width=0.5pt ] (5.5,12.25) rectangle (6.5,11);
      \draw [, line width=0.5pt ] (6.75,11) rectangle (7.75,9.75);
      \draw [, line width=0.5pt ] (8,9.75) rectangle (9,8.5);
      \draw [, line width=0.5pt ] (9.25,8.5) rectangle (11.25,4.75);
      \draw [, line width=0.5pt ] (11.5,8.5) rectangle (12.5,7.25);
      \draw [, line width=0.5pt ] (12.75,7.25) rectangle (13.75,6);
      \draw (14,6) rectangle (15,4.75);
      \draw [line width=1pt, short] (2.5,12.5) -- (2,12.5);
      \draw [line width=1pt, short] (2,12.5) -- (2,4.5);
      \draw [line width=1pt, short] (2,4.5) -- (2.5,4.5);
      \draw [line width=1pt, short] (14.75,4.5) -- (15.25,4.5);
      \draw [line width=1pt, short] (15.25,4.5) -- (15.25,12.5);
      \draw [line width=1pt, short] (15.25,12.5) -- (14.75,12.5);
      \draw [, line width=0.5pt ] (20,12.25) rectangle (30,11.25);
      \draw [, line width=0.5pt ] (20,11) rectangle (25,9.75);
      \draw [, line width=0.5pt ] (20,9.5) rectangle (22.5,9.25);
      \draw [, line width=0.5pt ] (20,9) rectangle (21.25,8.5);
      \draw [, line width=0.5pt ] (21.25,8.25) rectangle (22.5,7.75);
      \draw [, line width=0.5pt ] (22.5,7.5) rectangle (25,7.25);
      \draw [, line width=0.5pt ] (22.5,7) rectangle (23.75,6.5);
      \draw [, line width=0.5pt ] (23.75,6.25) rectangle (25,5.75);
      \draw [, line width=0.5pt ] (25,5.5) rectangle (30,4.25);
      \draw [, line width=0.5pt ] (25,4) rectangle (27.5,3.75);
      \draw [, line width=0.5pt ] (25,3.5) rectangle (26.25,3);
      \draw [, line width=0.5pt ] (26.25,2.75) rectangle (27.5,2.25);
      \draw [, line width=0.5pt ] (27.5,2) rectangle (30,1.75);
      \draw [, line width=0.5pt ] (27.5,1.5) rectangle (28.75,1);
      \draw [, line width=0.5pt ] (28.75,0.75) rectangle (30,0.25);
      \draw [line width=1pt, short] (20,12.5) -- (19.5,12.5);
      \draw [line width=1pt, short] (19.5,12.5) -- (19.5,0);
      \draw [line width=1pt, short] (19.5,0) -- (20,0);
      \draw [line width=1pt, short] (29.75,0) -- (30.25,0);
      \draw [line width=1pt, short] (30.25,0) -- (30.25,12.5);
      \draw [line width=1pt, short] (30.25,12.5) -- (29.75,12.5);
    \end{circuitikz}
  }
  \caption{%
    On the left a schematic depiction of a multi-stage stochastic matrix with three levels.
    On the right a schematic tree-fold matrix with 4 layers.
    All entries outside of the rectangles must be zero.
    Entries within rectangles can be non-zero.
  }
  \label{fig:treefoldmultistage}
\end{figure}

Because it is \NPh~\cite{Kar72}, we study tractable subclasses corresponding to block-structured matrices, namely so-called multi-stage stochastic and tree-fold matrices, depicted in Figure~\ref{fig:treefoldmultistage}.
Both classes have been a subject of much study and have many theoretical and practical applications; see the survey of Schultz et al.~\cite{SchultzSvdV96} as well as an exposition article by Gaven\v{c}iak et al.~\cite{GavenciakKK22} for examples of~applications of multi-stage stochastic matrices, and see~\cite{ChenMYZ17,DeLoera2008,GavenciakKK22,JansenKMR22,KnopK18,KnopKLMO21,KnopKM20,KnopKM20b} for examples of applications of~\ref{IP} with tree-fold matrices.

Equivalently, assuming $\|A\|_\infty$ is small, multi-stage stochastic matrices are exactly those with small primal treedepth $\td_P(A)$, and tree-fold matrices are those with small dual treedepth $\td_D(A)$~\cite[Lemmas 25 and 26]{KouteckyLO:2018}; we will formally introduce these notions later.
%These are defined as follows.
%For positive integers $m \leq n$ we set $[m,n] \df \{m,\ldots, n\}$ and $[n] \df [1,n]$.
%Let $G_P(A)$ denote the \emph{primal graph of $A$}, which has \([n]\) as its vertex set, and an edge between vertices $i$ and $j$ exists if $A$ contains a row which is nonzero in coordinates $i$ and $j$.
%The \emph{dual graph of $A$} is $G_D(A) \df G_P(A^{\intercal})$.
%The \emph{treedepth} of a graph~\(G\), denoted \(\td(G)\), is the smallest height of a rooted forest $F$ such that each edge of $G$ is between vertices which are in a descendant-ancestor relationship in $F$.
%The \emph{primal treedepth of $A$} is $\td_P(A) \df \td(G_P(A))$, and analogously the \emph{dual treedepth of $A$} is $\td_D(A) \df \td(G_D(A))$.
%Then, the sparsity measure $d$ is defined as $d \df \min \{\td_P(A), \td_D(A)\}$.
%induced by two natural parameters of the constraint matrix $A$: its numeric measure $a$ and its sparsity measure $d$ defined as follows.
%We also require that the largest coefficient in absolute value $\|A\|_\infty$ is small.
Note that both restrictions on $\|A\|_\infty$ and treedepth are necessary to obtain efficient algorithms:
\textsc{Independent Set} can be modelled as an~\ref{ILP} with~\(\norm{A}_\infty = 1\), and \textsc{Knapsack} can be modelled as an~\ref{ILP} with one row, meaning that both cases are \NPh.
% While one could argue that \textsc{Independent Set} is sparse in the sense that its straightorward~\ref{IP}~formulation has only two non\=/zeros per row, it is not possible to solve it quickly using our approach.
% The reason for that is that there are instances such that the Graver basis (see Definition~\ref{def:graver} later) of the associated~\ref{IP} has size~\(\Omega(2^n)\).
% A full binary tree would be one such example, see~\cite{JesusBook} for more details.

%Equivalently, the sparsity measure is small if $A$ is either a so-called multi-stage or tree-fold matrix with small block sizes, respectively for small $\td_P(A)$ and $\td_D(A)$~, that is, it looks roughly as follows (Figure~\ref{fig:treefoldmultistage}):

% Denote by $\la A, f, \veb, \vel, \veu \ra$ the binary encoding length of an~\ref{IP} instance.\footnote{We define the encoding length of $f$ to be the length of $f_{\gap}$, which is the difference between the maximum and minimum values of $f$ on the domain.}
%The function $f$ is given by an oracle.
%Note that already for $a=1$ or $d=1$~\ref{ILP} is \NPh \cite{E+22arxiv}.
%Moreover, arguably the two most important tractable classes of~\ref{IP} are formed by instances whose constraint matrix is either totally unimodular or has small number $n$ of columns, yet our results are incomparable with either: the class of totally unimodular matrices might have large $d$, but has $a=1$, and the matrices considered here have variable $n$.

A stream of papers~\cite{HemS03,AschenbrennerH07,CslovjecsekEHRW21,CslovjecsekEPVW21,DeLoera2008,EisenbrandHK18,HemmeckeOR13,JansenLR20,DBLP:journals/mp/Klein22} spanning the last 20+ years has brought gradually more and more general as well as faster algorithms for block-structured IPs, as well as provided complexity lower bounds~\cite{JansenKL23,KnopPW20}.
The pinnacle of these developments is the fundamental result that~\ref{IP} is solvable in time~$g(\|A\|_\infty,d) (n^\omega + \min(m, n)nm) \log \|\veu-\vel\|_\infty \log f_{\gap}$~\cite{MOR} where~$d = \min\{\td_P(A), \td_D(A)\}$, $f_{\gap} = \max_{\vex, \vey: \vel \leq \vex, \vey \leq \veu} f(\vex) - f(\vey)$, \(g\) is some computable function, and~\(n^\omega\) is time required to multiply two~\(n \times n\)~matrices.
Cslovjecsek et al.~\cite{CslovjecsekEHRW21,CslovjecsekEPVW21} have designed an algorithm for~\ref{ILP} with complexity $g(\|A\|_\infty,d) n \log^{2^{\bigO(d)}} n$ by coupling the multidimensional search technique~\cite{DBLP:journals/jal/NortonPT92} with new structural results.
This algorithm is \emph{strongly} near-linear time in the sense that the number of arithmetic operations it performs does not depend on the numbers in~$\veb, \vel, \veu, \vew$ on input.
While the exponent over $\log n$ could conceivably be reduced, we consider the complexity of~\ref{ILP} in the regime of small $\td_P(A)$ or $\td_D(A)$ and with small coefficients to be essentially settled by these results.

\subsection{Our Contribution}

The situation is quite different for~\ref{IP} with a general separable convex function: no strongly polynomial algorithm is possible for~\ref{IP} in general, in fact, at least $n \log \|\veu - \vel\|_\infty$ comparisons are required to solve an~\ref{IP} instance by the following information-theoretic argument.
For each $i \in [n]$, finding the integer minimum of a convex function $f_i: \R \to \R$ in the interval $[l_i, u_i]$ can be optimally done by binary search with $\log (u_i - l_i)$ comparisons (and cannot be done faster), and the integer minimum of $f(\vex) = \sum_i f_i(x_i)$ is simply the vector of minima for each coordinate.
Thus, minimizing $f(\vex)$ over the box $[\vel, \veu]$ takes at least~$n \log \|\veu - \vel\|_\infty$~comparisons.
It is then not surprising that the techniques of~\cite{CslovjecsekEHRW21,CslovjecsekEPVW21} are inherently limited to linear objectives and cannot be extended to the more general separable convex case.
We focus on this question:

\begin{quote}
	\emph{What is the complexity of~\ref{IP} with a separable convex objective given by a comparison oracle, parameterized by $\td_P(A)$ or $\td_D(A)$ and $\|A\|_\infty$?}
\end{quote}

In the primal treedepth case, we design an algorithm matching the lower bound, up to the dependency on the parameters $\td_P(A)$ and $\|A\|_\infty$.
\begin{theorem}
  \label{thm:primal-alg}
  There is a computable function $g$ and an algorithm which solves~\ref{IP} in time~\(g(\td_P(A), \|A\|_\infty) n \log(\max(\|\veu - \vel\|_\infty, \norm{\veb}_\infty))\).
\end{theorem}
Note that the dependence on~\(\log\norm{\veb}_\infty\) above (and below) is only necessary to find an initial feasible solution.
It is usually the case that when we model an instance of a problem with~\ref{IP}, we can also provide an initial feasible solution for the instance.
Moreover, a recent result of Cslovjecsek et al.~\cite{CslovjecsekKLPP24} shows that the feasibility problem can be solved in linear time (up to dependence on the parameters) for $n$-fold and 2-stage matrices.
Combining this result with the algorithm from Theorem~\ref{thm:primal-alg} results in an algorithm for~\ref{IP} with 2\=/stage constraint matrices whose running time can be bounded by~\(g'(\td_P(A), \|A\|_\infty) n \log\|\veu - \vel\|_\infty\) for some computable function~\(g'\).
% via Lemma~\ref{lem:initial-feasible-sol}.

In the dual treedepth case (i.e., tree-fold matrices), we devise an algorithm which misses the lower bound only by a $\log n$ factor:
\begin{theorem}
  \label{thm:dual-alg}
  There is a computable function $g$ and an algorithm which solves~\ref{IP} in time~\(g(\td_D(A), \norm{A}_\infty) n \log(n) \log(\max(\norm{\veu - \vel}_\infty, \norm{\veb}_\infty))\).
\end{theorem}
% We conjecture that the information-theoretic lower bound could be strengthened when $A$ has small dual treedepth, and that our algorithm is actually optimal.
% We provide some indirect evidence in this direction in Appendix~\ref{sec:hochbaum}.
For~\ref{IP} with \(n\)\=/fold constraint matrices we can obtain an algorithm whose running time does not depend on~\(\|\veb\|_\infty\), analogously to the case of 2\=/stage matrices:
First we use the algorithm of Cslovjecsek et al.~\cite{CslovjecsekKLPP24} to obtain an initial feasible solution which we then provide to the algorithm from Theorem~\ref{thm:dual-alg} and this results in an algorithm whose running time can be bounded by~\(g'(\td_D(A), \norm{A}_\infty) n \log(n) \log\norm{\veu - \vel}_\infty\) for some computable function~\(g'\).

Note that focusing on convex (non-linear) objective functions is well justified.
For example, Bertsimas et al. note in their work~\cite{BertsimasKM15} on the notorious subset selection problem in statistical learning that, over the past decades, algorithmic and hardware advances have elevated \emph{convex} (i.e., non-linear) optimization to a level of relevance in applications comparable to its linear counterpart.
%\byasaf{Sentence too long.}
Moreover, our Theorem~\ref{thm:dual-alg} directly speeds up algorithms relying on $n$-fold~\ref{IP} with separable convex objectives~\cite{KnopK18,GavenciakKK22,MIMO}.

\paragraph{Techniques.}
Both our algorithms are based on scaling, akin to the well-known capacity scaling algorithm for maximum flow~\cite[Section 7.3]{AhujaMO:1993} or the scaling algorithm for~\ref{IP} of Hochbaum and Shanthikumar~\cite{HS}.
Theorem~\ref{thm:primal-alg} utilizes a recent strong proximity result of Klein and Reuter~\cite{KleinR22}.
Theorem~\ref{thm:dual-alg} is much more delicate; to show it, we adopt an approach similar to iterative rounding used in approximation algorithms~\cite{lau2011iterative}.
This approach is enabled by a new sensitivity result (Theorem~\ref{thm:sensitivity}) showing that ``de-scaling'' one variable at a time will lead to a sparse update; a novel dynamic data structure allowing fast sparse updates then completes the picture.
%We will give a more detailed description of the used techniques in Section~\ref{sec:used-techniques}

Note that our focus here is explicitly \emph{not} on the parameter dependence since the dependence on $g$ in our algorithm stems from a source common to most existing algorithms for those classes of~\ref{IP}.
So any positive progress is likely to translate also to our algorithms.

\subsection{Related Work}
Unless stated otherwise, we omit the dependency on the parameters (i.e., $\|A\|_\infty$, block sizes, $\td_P(A)$, $\td_D(A)$, etc.).
Jansen et al.~\cite{JansenLR20} have designed the first near-linear time algorithm for $n$-fold~\ref{ILP}, a subclass of~\ref{ILP} with small $\td_D(A)$, with complexity $n \log^{\bigO(1)} n \log f_{\gap} \log \|\veu-\vel, \veb\|_\infty$.
Their algorithm only works for linear objectives, but could be modified to work also for separable convex functions.
Like Theorem~\ref{thm:dual-alg}, it is also based on the idea of sparse updates, but uses a less efficient and less general color-coding-based approach.
%\byasaf{``This paragraph should contain also something regarding the MOR+ORL papers resulting from ATIP''}
Jansen et al.~\cite{JansenLR20} were also the first to study sensitivity in the context of $n$-fold~\ref{ILP}.
Eisenbrand et al.~\cite{MOR} give a weakly-polynomial algorithm in the case of separable convex objectives, and, in another paper~\cite{reducibility}, show how to reduce the dependency on the objective function, if given the continuous optimum.
Hunkenschröder et al.~\cite{ipco2024-lower-bounds} show lower bounds on the parameter dependencies for algorithms for block-structured~\ref{ILP}, however, our focus here is on the polynomial dependency.
%Cslovjecsek et al.~\cite{CslovjecsekEHRW21,CslovjecsekEPVW21} utilize the 
Cslovjecsek et al.~\cite{CslovjecsekKLPP24} give novel algorithms for $n$-fold and 2-stage~\ref{ILP} which work even if $A$ contains large coefficients in the horizontal or vertical blocks, respectively.
In order to achieve this, they avoid the usual approach based on the Graver basis, but also seem inherently limited to the linear case; in the case of $2$-stage~\ref{ILP} even to the case of testing the feasibility, though this has been overcome by Eisenbrand and Rothvoss~\cite{EisenbrandR25}.

A major inspiration for our work is the seminal paper of Hochbaum and Shanthikumar~\cite{HS} which explores ideas such as proximity and scaling in the context of separable convex optimization.
Compared to them, we focus on the case of small treedepth while they study programs with small subdeterminants.
However, our iterative scaling approach, and the associated dynamic data structure, are new.
Finally, various dynamic data structures allowing for sparse updates are a major focus of recent breakthroughs regarding LP~\cite{DongGLSY24,DongLY21,ChenKLPGS22,near-linear-flows-2}.

%\medskip

%Eisenbrand et al.~\cite[Theorem 80]{monster} design an algorithm for $\td_D(A)$ with complexity $n \log n \log f_{\gap} \log \|\veu-\vel, \veb\|_\infty$

%Jansen - color coding - very weak, because it depends on both $\log \|\veu - \vel\|_\infty$ and $\log f_{\gap}$ (I think) - it just takes the old algorithm and does sparse updates using color coding; so we have $\log n$ instead of some $\poly\log n$ and we eliminate the $\log f_{\gap}$ thanks to scaling.

%IPCO '24 lower bounds - addresses the parameter dependence question

%Cslovjecsek et al. -- why it cannot be extended (for $n$-folds proximity relies on having a vertex solution)
%
%Our SODA '24 paper - avoids dependence on graver norms, but does not apply to sep convex
%
%Hochbaum-Shanthikumar scaling algo -- we use Graver bounds which are better
%
%Hochbaum - gives lower bound for the case of 1 constraint - this is somewhat counter-intuitive, that a lb with no constraints is stronger than a lb with one constraint - the constraint actually makes the problem easier.
% GAP - postpone discussion until after we have explained our algorithm
%
%proximity and sensitivity is a big topic
%
%dynamic data structures - STOC '21 paper about tw LPs (also explain why we can't use this + proximity)

%% file: preliminaries.tex
%! TEX root = main.tex
\section{Preliminaries} \label{sec:prel}
\sv{\toappendix{\section{Additional Material for Section~\ref{sec:prel}} \label{app:sec:prel}}}

If~$A$ is a matrix, $A_{i,j}$ denotes the $j$-th coordinate of the $i$-th row, $A_{i, \bullet}$ denotes the $i$-th row and $A_{\bullet, j}$ denotes the $j$-th column.
We use $\log \df \log_2$.
For~\(a, b \in \Z\) we denote by~\([a, b] = \{c \in \Z \mid a \le c \le b\}\), and~\([a] \df [1, a]\).
We extend this notation to vectors~\(\veu, \vev \in \Z^a\) with~\(a \in \N\) where~\([\veu, \vev] \df \{\vew \in \Z^a \mid u_i \le w_i \le v_i \, \forall i \in [a]\}\).
For multiple real vectors~\(\veu_1, \ldots, \veu_n\) and any~\(1 \le p \le +\infty\) we denote by~\(\norm{\veu_1, \ldots, \veu_n}_p\) the~\(\ell_p\)\=/norm of the vector obtained by concatenating vectors~\(\veu_1, \ldots, \veu_n\).

The \emph{relaxation} of~\ref{IP} is the problem
\begin{align}
  \label{relax}
  \tag{RP}
  \min & \left\{f(\vex) \mid \ A\vex = \veb, \ \vel \leq \vex \leq \veu, \ \vex \in \R^n \right\} \enspace .
\end{align}

For an~\ref{IP} instance $\II$, let $\Sol(\II) \df \{\vex \in \Z^n \mid A\vex=\veb, \, \vel \leq \vex \leq \veu\}$ denote the set of feasible solutions of $\II$.
We will use the notion of \emph{centering an~\ref{IP}~instance at a vector $\vev \in \Z^n$} which defines a new~\ref{IP} instance~\(\bar{\II} = (A, \bar\veb, \bar\vel, \bar\veu, \bar{f})\) where~\(\bar\veb = \veb - A\vev\), \(\bar\vel = \vel - \vev\), \(\bar\veu = \veu - \vev\), \(\bar{f}(\vex) = f(\vex - \vev)\).
Centering at~\(\vev\) can be viewed as a translation~\(\tau(\vex) = \vex + \vev\).
Then~\(\tau\) is a bijection between~\(\Sol(\II)\) and~\(\Sol(\bar\II)\), and~\(\bar{f}\) only translates its input which it passes to the oracle for~\(f\).
Thus given an optimum~\(\vev^*\) of~\(\bar\II\) we can obtain an optimum of~\(\II\) in time~\(\Oh(n)\).
If~\(\vev\) is a feasible solution, i.e.~we have~\(A\vev = \veb\) and~\(\vel \le \vev \le \veu\), then we can perform the centering in time~\(\Oh(n)\) as we can skip the matrix multiplication~\(A\vev\) and directly set~\(\bar{\veb} = \vezero\).
% \toappendix{\begin{lemma}[Equivalent centered instance] \label{lem:centered}
%   Let an~\ref{IP} instance $\II$ be given, and $\vev \in \Z^n$.
%   Define an~\ref{IP} instance $\bar{\II}=(A,\bar{f},\bar{\veb}, \bar{\vel}, \bar{\veu})$ by
%   $$\bar{\veb} \df \veb - A\vev, \, \bar{\vel} \df \vel - \vev, \, \bar{\veu} \df \veu - \vev, \, \bar{f}(\vex) \df f(\vex -\vev) \enspace .$$
%   The translation $\tau (\vex) = \vex + \vev$ is a bijection from $\Sol(\bar{\II})$ to $\Sol(\II)$.
%   Moreover, $\vex$ is an optimal solution of $\bar{\II}$ if and only if $\tau(\vex)$ is an optimal solution of $\II$.
%   If $\vev \in \Sol(\II)$, then $\bar{\veb} = \vezero$, and $\vezero$ is feasible for $\bar{\II}$.
% \end{lemma}}

We use the standard way to find an initial feasible solution to~\ref{IP} akin to Phase I of the simplex method.
We formalize this in the following lemma.

\begin{lemma}[Folklore, {see \cite[Lemma 51]{monster}}]
  \label{lem:initial-feasible-sol}
  Let~\(\II = (A, \veb, \vel, \veu, f)\) be an~\ref{IP}~instance with~\(\vezero \in [\vel, \veu]\) (one can simply translate the instance in order to achieve this).
  Define vectors~\(\vew', \vel', \veu' \in \Z^m\) as
  \begin{align*}
    w'_i &\df \sign(b_i), & l_i' &\df \min\{0, b_i\}, & u_i' &\df \max\{0, b_i\}, & \forall i \in [m] \, ,
  \end{align*}
  let~\(\bar{\vel} = (\vel, \vel')\), \(\bar{\veu} = (\veu, \veu')\), and~\(A_I = (A\,I) \in \Z^{m \times (n + m)}\) where~\(I\) here is the~\(m \times m\) identity matrix.
  The vector~\((\vezero, \veb)\) is a feasible solution for the instance
  \begin{equation}
    \label{eq:ai-feasible-ip}
    \tag{\(A_I\)-feasible IP}
    \min \{\vew'\vex' \mid A_I(\vex, \vex') = \veb, (\vel, \vel') \le (\vex, \vex') \le (\veu, \veu'), \, (\vex, \vex') \in \Z^{n + m}\} \enspace ,
  \end{equation}
  and this instance has an optimum of the form~\((\vex, \vezero)\) (of value zero) if and only if~\(\II\) is feasible, and then~\(\vex\) is a feasible solution of~\(\II\).
\end{lemma}
Note that in~\eqref{eq:ai-feasible-ip} we have~\(\norm{\bar{\vel}}_\infty \le \norm{\vel, \veb}_\infty\) and~\(\norm{\bar{\veu}}_\infty \le \norm{\veu, \veb}_\infty\).

\toappendix{
We will use the following standard property of separable convex functions.
\begin{proposition}
  [{Separable convex superadditivity~\cite[Lemma 3.3.1]{JesusBook}}]
  \label{prop:superadditivity}
  Let $f(\vex) = \sum_{i=1}^n f_i(x_i)$ be separable convex, let $\vex \in \R^n$, and let $\veg_1,\dots,\veg_k \in \R^n$ be vectors that belong to the same orthant in~\(\R^n\).
  %conformal to $\vex$.
  Then
  \begin{align}
    \label{eq:conv_ineq}
    f \left( \vex + \sum_{j=1}^k \alpha_j \veg_j \right) - f(\vex)
                &\geq \sum_{j=1}^k \alpha_j \left( f(\vex + \veg_j) - f(\vex) \right)
  \end{align}
  for arbitrary integers $\alpha_1,\dots,\alpha_k \in \N$.
\end{proposition}

The following proposition follows straightforwardly from Proposition~\ref{prop:superadditivity}.
\begin{proposition}
  \label{prop:prox-superadd}
  Let $\vex, \vey_1, \vey_2 \in \R^n$, $\vey_1, \vey_2$ be from the same orthant, and $f$ be a separable convex function.
  Then
  \begin{equation}
    f(\vex + \vey_1 + \vey_2) - f(\vex+\vey_1) \geq f(\vex+\vey_2) - f(\vex) \enspace .
  \end{equation}
\end{proposition}
\begin{proof}
  Apply Proposition~\ref{prop:superadditivity} with $\vex \df \vex$, $\veg_1 \df \vey_1$, $\veg_2 \df \vey_2$, and $\alpha_1, \alpha_2 \df 1$, to get
  \begin{equation}
    \begin{split}
      f(\vex + \vey_1 + \vey_2) - f(\vex) &\geq \left(f(\vex + \vey_1) - f(\vex)\right) + \left(f(\vex + \vey_2) - f(\vex)\right) \\
                                          &= f(\vex + \vey_1) + f(\vex + \vey_2) - 2f(\vex) \enspace .
    \end{split}
  \end{equation}
  Adding~\(f(\vex) - f(\vex + \vey_1)\) to both sides then yields the statement.
  \qed
\end{proof}
}

\paragraph{Graphs of~\(A\) and treedepth.}
By~\(G_P(A)\) we denote the \emph{primal graph of~\(A\)} whose vertex set is~\([n]\) (corresponding to the columns of $A$), and we connect two vertices~\(i \neq j \in [n]\) if there exists a row~\(k\) of~\(A\) such that~\(A_{k, i} \neq 0\) and~\(A_{k, j} \neq 0\).
We denote~\(G_D(A) \df G_P(A^T)\) the \emph{dual graph of~\(A\)}.
From this point on we always assume that $G_P(A)$ and $G_D(A)$ are connected, otherwise $A$ has (up to row and column permutations) a diagonal structure with some number $d \geq 2$ of blocks, and solving~\ref{IP} amounts to solving $d$ smaller~\ref{IP} instances independently.
Thus, the bounds we will show provide similar bounds for each such independent instance.
\begin{definition}[Treedepth]
\label{def:tree-depth}
The \emph{closure~$\cl(F)$} of an undirected rooted tree $F$ is the undirected graph obtained from $F$ by making every vertex adjacent to all of its ancestors.
For a rooted tree~\(F\) its \emph{height~\(\height(F)\)} is the maximum number of vertices on any root-leaf path.
The \emph{treedepth~\(\td(G)\)} of a connected graph~$G$ is the minimum height of a rooted tree~$F$ where~\(V(F) = V(G)\) such that $ G\subseteq \cl(F)$, namely~\(E(G) \subseteq E(\cl(F))\).
A \emph{$\td$-decomposition of $G$} is a rooted tree $F$ such that $G \subseteq \cl(F)$.
A $\td$-decomposition $F$ of $G$ is \emph{optimal} if $\height(F) = \td(G)$.
\end{definition}
Computing $\td(G)$ is \NPh, but can be done in time $2^{\td(G)^2} \cdot |V(G)|$~\cite{ReidlRVS:2014}.
%\begin{proposition}[{\cite{ReidlRVS:2014}}]\label{prop:tddecomposition}
%The treedepth $\td(G)$ of a graph $G$ with an optimal $\td$-decomposition $F$ can be computed in time $2^{\td(G)^2} \cdot |V(G)|$.
%\end{proposition}
We define the \emph{primal treedepth of $A$} to be $\td_P(A) \df \td(G_P(A))$ and the \emph{dual treedepth of $A$} to be $\td_D(A) \df \td(G_D(A))$.
We assume that an optimal $\td$-decomposition is given since the time required to find it is dominated by other terms.
%Moreover, in many applications a small $\td$-decomposition of $G_P(A)$ or $G_D(A)$ is clear from the way $A$ was constructed.
%By definition, a graph $G$ has at most $\td(G)^2 |V(G)|$ edges because the closure of each root-leaf path of a $\td$-decomposition of $G$ contains at most $\td(G)^2$ edges, and there are at most $|V(G)|$ leaves.
Constructing $G_P(A)$ or $G_D(A)$ can be done in linear time if $A$ is given in its sparse representation because there are at most $\td(G) \cdot |V(G)|$ edges.
Throughout we shall assume that $G_P(A)$ or $G_D(A)$ are given.

In the construction of our novel dynamic data structure, we utilize the parameter topological height, introduced by~\cite{MOR}, and which has been shown to play a crucial role in complexity estimates of~\ref{IP}~\cite{MOR,ipco2024-lower-bounds}.
\begin{definition}[Topological height]
  \label{def:topheight}
  A vertex of a rooted tree $F$ is \emph{degenerate} if it has exactly one child, and \emph{non-degenerate} otherwise (i.e., if it is a leaf or has at least two children).
  %Note that if the root has only one child then it is degenerate.
  The \emph{topological height of $F$}, denoted $\ttd(F)$, is the maximum number of non-degenerate vertices on any root-leaf path in $F$.
  Equivalently, $\ttd(F)$ is the height of $F$ after contracting each edge from a degenerate vertex to its unique child.
  Clearly, $\ttd(F) \leq \height(F)$.

  Now we define the level heights of $F$ which relate to lengths of paths between non-degenerate vertices.
  For a root-leaf path $P=(v_{b_0}, \dots, v_{b_1}, \dots, v_{b_2}, \dots, v_{b_e})$ with $e$ non-degenerate vertices $v_{b_1}, \dots, v_{b_e}$ (possibly~$v_{b_0} = v_{b_1}$), define $k_1(P) \df |\{v_{b_0}, \dots, v_{b_1}\}|$, $k_i(P) \df |\{v_{b_{i-1}}, \dots, v_{b_i}\}|-1$ for all $i \in [2,e]$, and $k_i(P) \df 0$ for all $i > e$.
  For each $i \in [\ttd(F)]$, define $k_i(F) \df \max_{P: \text{root-leaf path}} k_i(P)$.
  We call $k_1(F), \dots, k_{\ttd(F)}(F)$ the \emph{level heights of $F$}.
\end{definition}

We illustrate why we distinguish between treedepth and topological height on the specific example of \PN{High Multiplicity Makespan Minimization on Unrelated Machines} where the ``High Multiplicity'' refers to the compact encoding of the input.
In this problem, we have~\(n\)~jobs grouped into~\emph{\(d\) job types} which we want to schedule on~\(m\)~machines.
It is known~\cite{KnopK18} that this problem has an~\ref{IP}~formulation with a constraint matrix~\(A = \begin{pmatrix}
    I   & I   & \cdots & I \\
    p^1 & 0   & \cdots & 0 \\
    0   & p^2 & \cdots & 0 \\
        &     & \ddots &  \\
    0   & 0   & \cdots & p^m 
  \end{pmatrix}\),
where~\(I\) are~\(d \times d\) identity matrices and~\(p^i\)'s are row vectors where the~\(j\)\=/th entry is the processing time of the~\(j\)\=/th job type on the~\(i\)\=/th machine.
The dual graph~\(G_D(A)\) of this matrix has a~\(\td\)\=/decomposition which is a path of length~\(d\) with~\(m\)~leaves attached to one of the ends of the path.
This problem admits an~\FPT algorithm with parameter~\(d\)~\cite{KnopK18} but it is \NPh if~\(d\) is part of the input already for the case when the maximum processing time is~\(2\)~\cite{r-hm-cmax-hard}.
However, the~\(\td\)\=/decompositions in both cases (either with~\(d\) as a parameter or as input) are the same except for the length of the path.
That is, the~\(\td\)\=/decompositions will be identical if we contract vertices of degree~2.

% We illustrate why we distinguish treedepth and topological height on the following example.
% Consider two arbitrary constraint matrices~\(A_1\) and~\(A_2\) with~\(n\)~columns where~\(A_1\) has~\(m\)~rows and~\(A_2\) has~\(2m\)~rows.
% In general, the~\(\td\)\=/decomposition of~\(G_D(A_1)\) could be a~\(m\)\=/vertex path and the~\(\td\)\=/decomposition of~\(G_D(A_2)\) could be a~\(2m\)\=/vertex path.
% But for both these matrices, the topological height would be just~2.
% \todo{I am still not wowed by this example. -- Tung}

\begin{definition}[{Dual Block-structured Matrix, \cite[Definition 7]{MOR}}]\label{def:dual-decomp}
  Let $A \in \Z^{m \times n}$ and $F$ be a $\td$-decomposition of $G_D(A)$.
  We say that $A$ is \emph{dual block-structured along $F$} if either $\ttd(F) = 1$, or if $\ttd(F) > 1$ and the following holds.
  Let $v$ be the first non-degenerate vertex in $F$ on a path from the root, $r_1, \dots, r_d$ be the children of $v$, $F_i$ be the subtree of $F$ rooted in $r_i$, and $m_i \df |V(F_i)|$, for $i \in [d]$, and
  \begin{align}
    A=
    \left(\begin{array}{cccc}
        \bar{A}_1 & \bar{A}_2 & \cdots & \bar{A}_d\\
        A_1 &      &  &  \\
            & A_2      &  &   \\
            &      & \ddots  &   \\
            &      &         & A_d
    \end{array}\right) \enspace . \tag{dual-block-structure} \label{eq:dual-block-structure}
  \end{align}
  where $d \in \N$, and for all $i \in [d]$, $\bar{A}_i \in \Z^{k_1(F) \times n_i}$, and $A_i \in \Z^{m_i \times n^i}$, $n_i \in \N$, and $A_i$ is block-structured along $F_i$.
  Note that $\ttd(F_i) \leq \ttd(F)-1$, $\height(F_i) \leq \height(F)-k_1(F)$, for $i \in [d]$.
\end{definition}

%\begin{definition}[Block-structured Matrix]\label{def:primal-decomp}
%  \martin{I have a feeling that we only need the definition of dual block structured matrix (in Section~\ref{sec:dual-alg}) and this primal one can be removed. -Tung}
%Let $A \in \Z^{m \times n}$ and $F$ be a $\td$-decomposition of $G_P(A)$.
%We say that $A$ is \emph{block-structured along $F$} if either $\ttd(F) = 1$, or if $\ttd(F) > 1$ and the following holds.
%Let $v$ be the first non-degenerate vertex in $F$ on a path from the root, $r_1, \dots, r_d$ be the children of $v$, $F_i$ be the subtree of $F$ rooted in $r_i$, and $n_i \df |V(F_i)|$, for $i \in [d]$, and
%\begin{align}
%	A=
%	\left(\begin{array}{ccccc}
%		\bar{A}_1 & A_1     &  &\\
%		\vdots &      &  \ddots & \\
%		\bar{A}_d&      &   & A_d
%	\end{array}\right), \tag{block-structure} \label{eq:block-structure}
%\end{align}
%where, for $i \in [d]$, $\bar{A}_i \in \Z^{m_i \times k_1(F)}$ where $k_1(F)$ is the first level height of $F$ and $m_i \in \N$, $A_i \in \Z^{m_i \times n_i}$, and $A_i$ is block-structured along $F_i$.
%Note that $\ttd(F_i) \leq \ttd(F)-1$, $\height(F_i) \leq \height(F)-k_1(F)$, for $i \in [d]$.
%\end{definition}
%A similar definition applies for the block structure of $G_D(A)$, namely it is the block structure of $G_P(A^T)$.
Whenever $A$ and $F$ are given, we will assume throughout this paper that $A$ is block-structured along $F$, as justified by~\cite[Lemma 5]{MOR}.
% As mentioned earlier, systems with small treedepth can be purified in linear time:
% \tung{Possibly unnecessary.}
% \begin{proposition}[{Bounded treedepth purification~\cite[Theorem 1.2 ]{FominLSPW:2018}}]\label{prop:tw_pure}
% 	Given $A \in \Z^{m \times n}$, a $\td$-decomposition $F$ of $G_P(A)$ or $G_D(A)$, and $\veb \in \Z^m$, in time $\Oh(\height(F)^2 (n+m))$ one can either declare $A\vex=\veb$ infeasible, or return a pure equivalent subsystem $A'\vex = \veb'$.
% \end{proposition}

\paragraph{Graver Bases.}
Define a partial order $\sqsubseteq$ on $\R^n$ as follows:
for $\vex,\vey\in\R^n$, write $\vex\sqsubseteq \vey$ and say that $\vex$ is \emph{conformal} to $\vey$ if~\(\vex \le \vey\), and for each~$i \in [n]$, $x_iy_i\geq 0$ holds, i.e.~\(\vex\) and~\(\vey\) lie in the same orthant.
Let $\ker_{\Z}(A) = \{\vex \in \Z^n \mid A\vex = \vezero\}$.
%It is well known that every subset of $\Z^n$ has finitely many $\sqsubseteq$\=/minimal elements~\cite{Gordan}.

\begin{definition}[Graver basis~\cite{Graver:1975}]\label{def:graver}
The {\em Graver basis} of an integer $m\times n$ matrix $A$ is the finite set $\G(A)\subset\Z^n$ of $\sqsubseteq$-minimal elements in $\ker_{\Z}(A) \setminus \{\vezero\}$.
\end{definition}

For a matrix~\(A \in \Z^{m \times n}\) and~\(1 \le p \le \infty\)  we denote~\(g_p(A) \df \max_{\veg \in \G(A)}\norm{\veg}_p\).
One important property of $\G(A)$ is as follows:

\begin{proposition}[{Positive Sum Property~\cite[Lemma 3.4]{Onn}}] \label{prop:possum}
Let $A \in \Z^{m \times n}$.
For any $\vex \in \ker_{\Z}(A)$, there exists an $n' \leq 2n-2$ and a decomposition $\vex = \sum_{j=1}^{n'} \lambda_j \veg_j$ with $\lambda_j \in \N$ and $\veg_j \in \G(A)$ for each $j \in [n']$, and with $\veg_j \sqsubseteq \vex$, i.e., all $\veg_j$ belonging to the same orthant as $\vex$.
\end{proposition}

%% file: scaling-algorithm.tex
\section{Scaling Algorithm \& Proximity Bounds}
\label{sec:scaling-alg}
\sv{\toappendix{\section{Additional Material for Section~\ref{sec:scaling-alg}} \label{app:sec:scaling-alg}}}

Our goal now is to show a scaling algorithm for~\ref{IP} with the following properties:
\begin{theorem}
  \label{thm:scaling-algo}
  Let~\(\mathcal{I} = (A, \veb, \vel, \veu, f)\) be an~\ref{IP}~instance, and~\(\beta = \ceil{\log\norm{\vel, \veu}_\infty} + 1\).
  There exists~\({\rho \in \R^+}\) which only depends on~\(A\) such that, given an initial feasible solution $\vex_0$ of~\(\mathcal{I}\), we can find the optimum of~\(\mathcal{I}\) by solving~\(\beta\)~\ref{IP}~instances~\((\mathcal{I}^i = (A, \vezero, \vel^i, \veu^i, f^i))_{i = 1}^{\beta}\), where, for every~\(i \in [\beta]\), \(f^i\) is a separable convex function if~\(f\) was, and~\(\norm{\veu^i - \vel^i}_\infty \le 3\rho\).
\end{theorem}

Note that in the statement of Theorem~\ref{thm:scaling-algo}, if for each of the instances~\(\II_i\) we have~\(\vel^i \le \vezero \le \veu^i\), then~\(\vezero \in \Sol(\II_i)\) due to the zero right hand side.

Let us start by giving the main ideas behind the algorithm.
Let~\(\II\) be an~\ref{IP}~instance. % for which we want to find an optimum.
%\byasaf{Consider deleting the rest of this paragraph.}
For the rest of this section, we shall assume~\(\vezero \in \Sol(\II)\):
This is without loss of generality as we can obtain a feasible solution~\(\vez\) by the standard approach (Lemma~\ref{lem:initial-feasible-sol}) and then center~\(\II\) at~\(\vez\).
The instance~\ref{eq:ai-feasible-ip} has small $\td_P(A)$ or $\td_D(A)$ if the original instance did, and has the same $\|A\|_\infty$; this will be used for establishing the algorithms in Theorems~\ref{thm:primal-alg} and~\ref{thm:dual-alg}.

For~\(s \in \N\), let the \emph{\(s\)\=/scaled lattice} be~\(s\Z^n \df \{s\vea \mid \vea \in \Z^n\}\).
For any~\ref{IP} instance~\(\JJ = (A', \veb', \vel', \veu', f')\) and any~\(s \in \N\), let its~\emph{\(s\)\=/scaled instance} be
\begin{align}
  \label{eq:s-ip}
  \tag{s-IP}
  \min & \left\{f(\vex) \mid \ A'\vex = \veb', \ \vel' \leq \vex \leq \veu', \ \vex \in s\Z^n \right\} \enspace ,
\end{align}
where~\((A', \veb', \vel', \veu', f', s)\) is the input.
The only difference between~\ref{IP} and~\ref{eq:s-ip} is that in the former we look for a solution in~\(\Z^n\) while in the latter in~\(s\Z^n\).
For notational convenience, we will also define a ``reverse version'' of~\ref{eq:s-ip}, called~\ref{eq:ip-scaled}, where instead of searching on a coarser lattice, we shrink the feasible region, but ``upscale'' each point with respect to the objective function~\(f\):
\begin{align}
  \label{eq:ip-scaled} \tag{IP-scaled}
  \min & \left\{ f(s \vex^\prime ) \mid \ A\vex^\prime = \veb, \ \lceil\tfrac{1}{s}\rceil\vel \leq \vex^\prime \leq \lfloor\tfrac{1}{s}\rfloor\veu, \ \vex^\prime \in \Z^n \right\}.
\end{align}
where~\((A, \veb, \vel, \veu, f, s)\) is the input.
For~\(s \in \N\) let~\(\II_s\) be the~\(2^s\)\=/scaled instance of~\(\II\).
Observe that~\(\vezero\) is the only feasible solution in~\(\II_\beta\) ,thus it is the optimum of~\(\II_\beta\).
In~\(\II_{\beta - 1}\) it still holds that \(\vezero\) is a feasible solution, so we may use it as an initial feasible solution from which we search for an optimum~\(\vex^{\beta - 1}\) of~\(\II_{\beta - 1}\), and so on for $\beta-2$ etc.
That is, we let~\(\vex^{\beta} = 0\), and then, for every~\(i = \beta - 1, \beta - 2, \ldots, 0\), we search for an optimum~\(\vex^{i}\) of~\(\II_i\) with~\(\vex^{i + 1}\) as an initial feasible solution, and we center~\(\II_i\) at~\(\vex^i\).
We prove that for each instance~\(\II_i\), there exists an optimum $\vex^i$ near the initial solution $\vex^{i+1}$, that is, with~\(\norm{\vex^{i + 1} - \vex^i}_\infty\) small.
So, we can restrict the original bounds to this smaller box, corresponding to ~\(\norm{\veu - \vel}_\infty\) being small.
Each~\(f^i\) is simply a translation and scaling of $f$, so it is separable convex.
%In order to do so, we leverage proximity bounds which we introduce in the following subsection.

\subsection{A Scaling Proximity Bound}
\label{sec:scaling-alg:proximity}

A \emph{proximity bound} in optimization bounds the distance between optima of related formulations of the same problem, e.g., an integer program and its relaxation, etc.~\cite{HS,HKW,CslovjecsekEHRW21,KnopKLMO2023,MurotaT04}.
The notion of proximity bound we use is the following.
\begin{definition}[Conformal proximity bound, \cite{CslovjecsekEPVW21}]
  \label{def:proximity}
  Let \(1 \leq p \leq +\infty\).
  We say that \(\PP_p(A) \geq 0\) is the \emph{conformal \(\ell_p\)-proximity bound of \(A\)} if it is the infimum of reals \(\rho \geq 0\) satisfying the following: for every~\ref{IP} with \(A\) as its constraint matrix, for every fractional solution \(\vex\) of~\ref{relax} and every integer solution \(\vez\) of~\ref{IP}, there is an integer solution \(\vez'\) of~\ref{IP} such that~\(\norm{\vex - \vez'}_p \leq \rho\) and~\(\vez' \sqsubseteq \vex - \vez\).
  % \begin{align*}
  %   \qquad &\text{and} \qquad \enspace .
  % \end{align*}
\end{definition}

An advantage of Definition~\ref{def:proximity} compared to prior notions of proximity bounds is that the bound~\(\rho\) does not depend on~\(f\).
Our main result here is:
%With Definition~\ref{def:proximity}, we can state our scaling proximity bound:

\begin{reptheorem}{thm:scaling-proximity}
  Let~\(1 \le p \le +\infty\), $\II$ be an~\ref{IP} instance,~\(\vez^*\) be an optimum of~\(\II\), and $\II'$ be~\(s\)\=/scaled~\(\II\) with~\(s \in \N\).
  Then there exists an optimum~\(\hat{\vez}\) of~\(\II'\) with
  \begin{equation}
    \label{eq:scaling-proximity}
    \|\vez^* - \hat{\vez}\|_p \le (s + 1)\PP_p(A),
  \end{equation}
  and for every optimum~\(\hat{\vez}\) of~\(\II'\) there exists an optimum~\(\vez^*\) of~\(\II\) satisfying~\eqref{eq:scaling-proximity}.
\end{reptheorem}

% where~\((A, \veb, \vel, \veu, f, s)\) is the input.
% Since the solution of~\ref{eq:s-ip} is required to lie in the~\(s\Z^n\), \eqref{eq:s-ip-zero} is effectively equivalent to a ``scaled \ref{IP}'' whose bounds are shrunk by a factor of $\tfrac{1}{s}$ when compared to the original instance:

We need the following lemma which shows that, for separable convex functions, a conformal proximity bound is indeed useful to relate optima of~\ref{IP} and~\ref{relax}.
%that for every optimum~\(\vez\) of an~\ref{IP} there exists an optimum~\(\vex\) of its relaxation~\ref{relax} such that~\(\|\vez - \vex\|_p \le \PP_p(A)\) for any~\(1 \le p \le +\infty\) and vice-versa.
The proof is a re-phrasing of known arguments~\cite{HS,HKW,JesusBook}.

\begin{figure}
  \centering
  \includegraphics[width=0.75\textwidth]{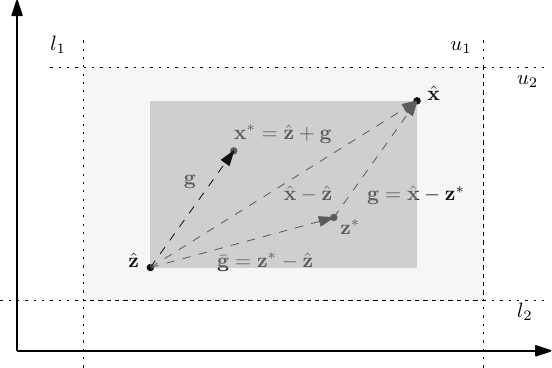}
  \caption{%
    The situation of Lemma~\ref{lem:prox-optima}: the feasible region between lower and upper bounds is the light grey rectangle; the dark grey rectangle marks the region of vectors which, when translated to $\hat{\vez}$, are conformal to $\hat{\vex} - \hat{\vez}$.
    The picture makes it clear that $\veg$ and $\bar{\veg}$ are conformal to $\hat{\vex} - \hat{\vez}$, and that the identity $\hat{\vex} = \hat{\vez} + \veg + \bar{\veg}$ holds.
  }
  \label{fig:prox-optima}
\end{figure}

\begin{lemma}
  \label{lem:prox-optima}
  Let~\(1 \le p \le +\infty\), \(\hat{\vez}\) be an optimum of~\ref{IP}, and~\(\hat{\vex}\) an optimum of the relaxation~\ref{relax} of~\ref{IP}.
  Then there exist~\(\vex^* \in \R^n\) and~\(\vez^* \in \Z^n\) optima of~\ref{relax} and~\ref{IP}, respectively, with $\norm{\hat{\vex} - \vez^*}_p = \norm{\vex^* - \hat{\vez}}_p \le \PP_p(A)$.
\end{lemma}
\begin{proof}
  See Figure~\ref{fig:prox-optima} for an illustration of the proof.
  By Definition~\ref{def:proximity}, there exists a $\vez^* \in \Z^n$ with $\vez^* \sqsubseteq \hat{\vex} - \hat{\vez}$ and $\|\vez^* - \hat{\vex}\|_p \leq \PP_p(A)$.
  Let $\veg =  \hat{\vex} - \vez^*$ and $\bar{\veg} = \vez^* - \hat{\vez}$ and note that $\hat{\vex} = \hat{\vez} + \veg + \bar{\veg}$, and~\(\vez^* = \hat{\vez} + \bar{\veg} = \hat{\vex} - \veg\).
  Now define~\(\vex^* \df \hat{\vez} + \veg\).
  By the fact that both $\hat{\vez}$ and $\hat{\vex}$ lie within the bounds $\vel$ and $\veu$ and that both $\veg$ and $\bar{\veg}$ are conformal to $\hat{\vex} - \hat{\vez}$, we see that both $\vex^*$ and $\vez^*$ also lie within the bounds $\vel$ and $\veu$.
  Thus $\vex^*$ is a feasible solution of~\ref{relax} and $\vez^*$ is a feasible solution of~\ref{IP}.
  We can also write
  \begin{equation}
    \hat{\vex} - \hat{\vez} = (\vex^* - \hat{\vez}) + (\vez^* - \hat{\vez}),
  \end{equation}
  which using Proposition~\ref{prop:prox-superadd} with values $\vex = \hat{\vez}$, $\vey_1 = \veg$, $\vey_2 = \bar{\veg}$, gives
  \begin{equation}
    f(\hat{\vex}) - f(\vex^*) \geq f(\vez^*) - f(\hat{\vez}) \enspace .
  \end{equation}
  Since $\hat{\vex}$ is a continuous optimum and $\vex^*$ is a feasible solution to~\ref{relax}, the left hand side is non-positive, and so is $f(\vez^*) - f(\hat{\vez})$.
  But since $\hat{\vez}$ is an integer optimum it must be that $\vez^*$ is another integer optimum and thus $f(\hat{\vez}) = f(\vez^*)$, and subsequently $f(\vex^*) = f(\hat{\vex})$ and thus $\vex^*$ is another continuous optimum.
  \qed
\end{proof}

\begin{proof}[of Theorem~\ref{thm:scaling-proximity}]
  We will show one direction, as the other direction is symmetric.
  Let $\vez^*$ be an optimum of~\ref{IP}.
  By Definition~\ref{def:proximity} and Lemma~\ref{lem:prox-optima}, there exists an optimum $\vex^*$ to the continuous relaxation~\ref{relax} with
  \begin{align}
    \label{eq:aux2}
    \norm{\vez^* - \vex^*}_p &\leq \PP_p(A) \enspace .
  \end{align}
  As the continuous relaxations coincide, $\vex^*$ is also an optimum to the continuous relaxation of~\ref{eq:ip-scaled}.
  Substituting $\vex^\prime \df \tfrac{1}{s} \vex$, 
  we obtain an objective-value preserving bijection between the solutions of the continuous relaxation of~\ref{IP} and the solutions of the continuous relaxation of~\ref{eq:ip-scaled}.
  In particular, $\bar{\vex} \df \tfrac{1}{s} \vex^*$ is an optimum to the continuous relaxation of~\ref{eq:ip-scaled}.

  Again by proximity (Lemma~\ref{lem:prox-optima}), the instance~\ref{eq:ip-scaled} has an optimal solution $\bar{\vez}$ with 
  \begin{alignat}{3}
    \allowdisplaybreaks
    \label{eq:aux3}
    \norm{\bar{\vex} - \bar{\vez}}_p &\leq \PP_p(A) & \quad &\Leftrightarrow \quad & \norm{s\bar{\vex} - s\bar{\vez}}_p &\leq s \cdot \PP_p(A) \enspace .
  \end{alignat}
  Substituting back, $\hat{\vez} \df s \bar{\vez}$ is an optimal solution to~\ref{eq:s-ip}.
  The claim follows by triangle inequality: $\norm{\vez^* - \hat{\vez}}_p \leq \norm{\vez^* - \vex^*}_p + \norm{\vex^* - \hat{\vez}}_p \leq (s+1) \PP_p(A)$.
  \qed
\end{proof}

\subsection{Scaling Algorithm for~\ref{IP}}

As a $2^{k}$-scaled instance is also a $2$-scaled instance of a $2^{k-1}$-scaled instance, we get the following algorithm to solve a general~\ref{IP}.
The algorithm is parameterized by some upper bound $\rho$ on the proximity bound $\PP_\infty(A)$ (see Theorem~\ref{thm:scaling-proximity}).
\begin{enumerate}[label=(\arabic*)]
  \item
    Given an initial feasible solution $\vex_{0}$, recenter the instance at $\vex_0$, obtaining~\ref{IP} where $\vezero$ is feasible.
    Set~\(\vex^\beta = \vezero\).
  \item
    \label{enum:scaling-algo:scaling-step}
    For each~\(i = \beta - 1, \beta - 2, \ldots, 0\), perform the following steps:
    Intersect the box constraints of the $2^{i+1}$-scaled instance with the proximity bounds $\norm{\vex^{i + 1} - \vex}_\infty \leq 3\rho$.
    Solve this instance with initial solution $\vezero = \vex^{i + 1}$ to optimality, and let $\vex^{i}$ denote the optimum.
    Center the current instance at~\(\vex^{i}\).
  \item Output $\vex^0$.
\end{enumerate}

The~\(3\rho\)~bound in step~\ref{enum:scaling-algo:scaling-step} follows from the fact that we are viewing the~$2^k$\=/scaled instance as an~$s$\=/scaled instance of the~$2^{k-1}$\=/scaled instance where~\({s = 2}\), and using~$3 = s+1$ in the bound of Theorem~\ref{thm:scaling-proximity}.
%Analyzing the running time, we prove Theorem~\ref{thm:scaling-algo}.
\begin{proof}[of Theorem~\ref{thm:scaling-algo}]
  Given an initial feasible solution $\vex_0$, we center the original instance at $\vex_0$ which transforms its right hand side to $\veb - A \vex_0 = \vezero$ as required.
  It is clear that we need at most $\ceil{\log\norm{\vel, \veu}_\infty} + 1$ iterations, in each of which we apply Theorem~\ref{thm:scaling-proximity} with $s=2$ to the bounds $\vel^i, \veu^i$.
  \qed
\end{proof}

\begin{remark} \label{remark}
  We can somewhat decrease the number of iterations of the scaling algorithm as follows.
  In each instance~\(\II^i\) we have~\(\norm{\veu^i - \vel^i} \le 3\rho\).
  In the first~\(\ceil{\log\rho}\)~instances~\(\II^i\), we are searching in the~\(2^{\beta - i}\)\=/lattice.
  Let us view these instances as~\ref{eq:ip-scaled}~instances:
  then each has~\(\vel^i = 2^{-(\beta - i)}\vel\) and~\(\veu^i = 2^{-(\beta - i)}\veu\).
  But then for each of them we already have~\(\norm{\vel^i, \veu^i}_\infty \le \rho\), so the guaranteed bound on~\(\norm{\veu - \vel}_\infty\) is satisfied even without invoking the proximity theorem.
  Thus, we can skip the first~\(\ceil{\log \rho}\)~iterations by initially setting~\(k = \beta - \ceil{\log\rho}\), reducing the number of iterations to~\((\ceil{\log\norm{\vel, \veu}_\infty} + 1) / \rho\).
  This improved number of iterations matches the $\log(B/n)$ bound on the number of scaling phases of the algorithm of Hochbaum~\cite{Hochbaum94} whose parameter~\(B\) is $\|\veu-\vel\|_\infty$, and her proximity bound is $\Oh(n)$.
\end{remark}

%% file: primal-algorithm.tex
% TEX root = main.tex
\section{Primal Algorithm}
\label{sec:primal-alg}

Theorem~\ref{thm:primal-alg} is proven by instantiating the scaling algorithm of Theorem~\ref{thm:scaling-algo} with a new strong proximity of Klein and Reuter~\cite{KleinR22}, and the standard branching algorithm for each subinstance~\cite[Lemma 8]{MOR}.
%which is efficient when $A$ has small $\td_P(A)$ and $\|A\|_\infty$, e.g. when it is a $2$-stage or a multi-stage stochastic matrix with small coefficients.

%\tung{Repeat Theorem 1.}

The parameter dependency of the function $g$ is essentially determined by $g_\infty(A)$ and the proximity bound $\PP_\infty(A)$.
The currently best known bounds for both are triple-exponential in terms of $\td_P(A)$~\cite[Corollary 2.1, Lemma 2.3]{KleinR22}, and this seems to be optimal for $g_\infty(A)$, see~\cite[Theorem 1]{ipco2024-lower-bounds}.
Since the exact dependence of $g$ on $\td_P(A)$ and $\|A\|_\infty$ is not the focus of this work, we shall simply focus on proving the theorem in its stated form.
Klein and Reuter~\cite{KleinR22} (improving on Cslovjecsek et al.~\cite{CslovjecsekEPVW21}) give a bound on $\PP_\infty(A)$ independent of $n$:
\begin{proposition}[{\cite[Lemma 2.3]{KleinR22}}] \label{prop:csl}
There is a computable function $g'$ such that
\begin{equation}
  \PP_\infty(A) \leq g'(\td_P(A), \|A\|_\infty) \enspace .
\end{equation}
% https://drops.dagstuhl.de/opus/volltexte/2021/14614/pdf/LIPIcs-ESA-2021-33.pdf
\end{proposition}
\begin{proof}[of Theorem~\ref{thm:primal-alg}]
  We instantiate the scaling algorithm from Theorem~\ref{thm:primal-alg} as follows.
  We provide Proposition~\ref{prop:csl} as an upper bound~\(\rho\) on~\(\PP_\infty(A)\).
  We solve each of the~${\ceil{\log(\norm{\vel, \veu}_\infty)} + 1}$~\ref{IP}~instances in the scaling algorithm using the algorithm of Eisenbrand et al.~\cite[Lemma 8]{MOR} which has running time~\(\td_P(A)^2 (2\rho + 1)^{\td_P(A)}n\).
  Taking $g(\td_P(A), \|A\|_\infty) = \td_P(A)^2 (2\rho + 1)^{\td_P(A)}$ and the fact that $\rho =  g'(\td_P(A), \|A\|_\infty)$ for some computable function $g'$ concludes the proof.
\qed
\end{proof}

%We remark that the dependence on~\(\log\norm{\veb}_\infty\) in the running time is only necessary to find an initial feasible solution via Lemma~\ref{lem:initial-feasible-sol}.

%% file: sensitivity.tex
% TEX root = ./main.tex
\section{Scaling and Sensitivity}
\label{sec:sensitivity}
\sv{\toappendix{\section{Additional Material for Section~\ref{sec:sensitivity} \label{app:sec:sensitivity}}}}

The situation is more delicate regarding the algorithm for dual treedepth (Theorem~\ref{thm:dual-alg}).
Not only is there no analogue of the proximity result of Klein and Reuter~\cite{KleinR22}, in fact, Cslovjecsek et al.~\cite[Proposition 4.1]{CslovjecsekEHRW21} show that there are instances of~\ref{IP} with small~\(\td_D(A) + \norm{A}_\infty\) where an integer optimum is~\(\Omega(n)\) far in the~\(\ell_\infty\)\=/norm from any continuous optimum.
The proximity results for stronger relaxations~\cite{CslovjecsekEHRW21,KnopKLMO23} are also of no obvious help.
Their bounds which are independent of~\(n\) only apply to vertex solutions, which are guaranteed to include an optimum only for linear objectives.
Moreover, solving these stronger relaxations is either not near-linear time (reliant on the ellipsoid method~\cite{KnopKLMO23}) or limited to linear objectives~\cite{CslovjecsekEHRW21}.

Let us re-examine the scaling algorithm.
In each iteration, we go from an optimum in the lattice~\(2s\Z^n\) to an optimum in~\(s\Z^n\).
From the point of view of~\ref{eq:ip-scaled} considering~\(s\Z^n\) instead of~\(2s\Z^n\) means that for every~\(i \in [n]\) we modify the separable objective function from~\(f_i(2sx_i')\) to~\(f_i(sx_i')\), the lower bound from~\(\frac{1}{2s}l_i\) to~\(\frac{1}{s}l_i\), and the upper bound from~\(\frac{1}{2s}u_i\) to~\(\frac{1}{s}u_i\).
Afterwards, we solve the new instance in the finer lattice~\(s\Z^n\) with an initial feasible solution which was the optimum in~\(2s\Z^n\).

Our approach for dual treedepth is to find the optimum solution in~\(s\Z^n\) by changing the scaling factor one variable at a time.
This means that once we modify~\(f_i, l_i, u_i\) for some~\(i \in [n]\), we immediately solve the instance resulting from refining coordinate $i$.
It is not clear why this should be advantageous -- after all, in each iteration of the scaling algorithm we need to solve~\ref{eq:ip-scaled}~\(n\)~times instead of just once.
The advantage lies in the fact that whenever we refine a single variable, %by modifying~\(f_i, l_i\), and~\(u_i\), 
then an optimum of the refined instance is close (in terms of the $\ell_1$-norm) to an optimum of the instance before this last refinement.
Such proximity bound will be our first contribution towards the proof of Theorem~\ref{thm:dual-alg}.

To show this proximity bound, we proceed in two steps.
First, we show a sensitivity theorem for changes in the box constraints $\vel, \veu$.
Then, we would like to use this theorem to show that refining few variables means only a small change in the optimal solution; however, this is not possible directly, because going from $l_i / 2$ to $l_i$ may be a big change in the absolute sense.
Thus, we use an encoding trick to circumvent this issue.
Let us start with the first step.

\begin{theorem}
  \label{thm:basic-sensitivity}
  Let~$\vex$ be an optimum of~\ref{IP}, $\vedelta \in \Z^n$, and $\vel' \df \vel + \vedelta$.
  If the~\ref{IP} with~$\vel$ replaced by $\vel'$ is feasible, then it has an optimum $\vex'$ with $\|\vex - \vex'\|_1 \leq \|\vedelta\|_1 g_1(A)$.
\end{theorem}
\begin{proof}
We will first prove the claim for the special case when $\vedelta$ is either the $i$-th unit vector $\vece_i$ or its negation $-\vece_i$ for some $i \in [n]$.
The theorem then follows by a repeated application of this claim and triangle inequality.
Specifically, we will start moving from $\vel$ to $\vel'$ by gradually decreasing all the coordinates in which $\vel'$ is smaller than $\vel$.
After this step, we have a vector $\vel''$ which is a coordinate-wise minimum of $\vel$ and $\vel'$.
The vector $\vex$ is feasible for all of the intermediate instances with lower bounds between $\vel$ and $\vel''$, and $\vex'$ is feasible the instance with lower bound~$\vel''$.
Next, we gradually increase all the coordinates in which $\vel'$ is greater than $\vel$, resulting in $\vel'$.
We have that $\vex'$ is feasible for all of the intermediate steps of the second phase.
This way we see that all the intermediate instances are feasible and thus also have some optimal solution, hence the claim applies to them.

Assume~$\vedelta = \pm \vece_i$ for some $i \in [n]$, and let us denote the original \ref{IP}~instance by~\(\II\) and the instance with~\(\vel' = \vel + \ve{\delta}\) by~\(\II'\).
Let $\vex'$ be an optimum of~\(\II'\) which is closest to $\vex$ (an optimum of~\(\II\)) in $\ell_1$-norm.
% If~\(\vex' = \vex\), then this theorem is trivially true so let us focus on the case when~\(\vex' \neq \vex\).
Since $\vex - \vex' \in \ker_{\Z}(A)$, it has a conformal decomposition $\vex - \vex' = \sum_j \veg_j$ with $\veg_j \in \G(A)$, $\veg_j \sqsubseteq \vex - \vex'$ for every $j$, and some $\veg_j$ potentially appearing with repetitions (see Proposition~\ref{prop:possum}).
% Let us fix a conformal decomposition of~\(\vex' - \vex\) and denote this multiset by~\(\ve{G}\).
We claim that there can actually be at most one element in the decomposition.

% We want to prove that each vector~\(\veg \in \ve{G}\) has~\(i \in \suppo(\veg)\).
% Suppose that~\(\OPT(\II) = \OPT(\II')\).
% If~\(\vex\) is feasible for~\(\II'\), then~\(\vex' = \vex\) and the theorem is trivially true.

% Let us distinguish the following four cases depending on whether:
% \begin{itemize}
%   \item \(\vex\) is feasible for~\(\II'\), and
%   \item whether~\(\OPT(\II) \neq \OPT(\II')\).
% \end{itemize}

% Let~\(\ve{G}^i = \{\vev \in \ve{G} \mid i \not\in \suppo(\veg)\}\) an let~\(\veg\) be the sum of vectors of~\(\ve{G}^i\).
% We want to prove that~\(\ve{G}^i\) is empty, i.e.~for each vector~\(\vev \in \ve{G}\) we have~\(i \in \suppo(\vev)\).
% So for contradiction, let us suppose that it is not.
% Due to the conformity of~\(\ve{G}\), we have~\(\veg \neq \vezero\) \martin{Check please!}.
% Let~\(\veg' = (\vex' - \vex) - \veg\), and let us inspect~\(\hat{\vex} = \vex + \veg'\).
% As~\(\vedelta = \pm \vece_i\) and~\(i \not\in \suppo(\veg)\), this means that~\(\hat{\vex}\) is feasible for both
% Only~\(l_i\) in~\(\vel\) has been modified and~\(i \not\in \suppo(\veg)\) so~\(\hat{\vex}\) is feasible for both~\(\II\) and~\(\II'\).
% This 

For contradiction assume that there are two or more elements in the conformal decomposition of~\(\vex' - \vex\).
Our plan is to use a proof analogous to that of Lemma~\ref{lem:prox-optima}.
We will define a vector $\veg \sqsubseteq \vex - \vex'$, use it to define $\hat{\vex} \df \vex' + \veg$ and $\tilde{\vex} \df \vex - \veg$, and show that $\tilde{\vex}$ is feasible for $\II'$ and $\hat{\vex}$ is feasible for $\II$.
This situation is a direct analogue of Figure~\ref{fig:prox-optima}, and the rest of the proof will be as well.

Let us fix some conformal decomposition of~\(\vex - \vex'\) and denote the multiset by~\(D\).
If there is no element~\(\veg \in D\) such that~\(i \in \suppo(\veg)\), then~\(\vex' \in [\vel, \veu]\) which in turn means that~\(\vex\) is also feasible and in fact optimum for~\(\II'\).
Hence, by the choice of~\(\vex'\), we have~\(\vex' = \vex\) and the bound on their distance follows trivially.

Now let us consider the case when there is at least one element in~\(\veg_j \in D\) with~\(i \in \suppo(\veg_j)\).
Set~$\veg = \veg_j$, \(\bar{\veg} \df (\vex - \vex') - \veg\), and accordingly $\hat{\vex} \df \vex' + \veg$ and $\tilde{\vex} \df \vex - \veg$.
We claim that~\(\hat{\vex}\) is still feasible for~\(\II\) and~\(\tilde{\vex}\) is still feasible for~\(\II'\).
In the case that~\(\veg_j\) is the only element of~\(D\) with~\(i \in \suppo(\veg_j)\), then as~\(\bar{\veg} \neq \vezero\) and~\(i \not\in \suppo(\bar{\veg})\), we have~\(\bar{\veg} \sqsubseteq \vex - \vex'\) which finishes the claim.
Otherwise assume that there is another element of~\(\veg_k \in D\) aside from~\(\veg_j\) with~\(i \in \suppo(\veg_k)\).
If $\vex'$ is infeasible for $\II$ or $\vex$ is infeasible for $\II'$, then it only violates the respective bound by $1$, and only in the $i$-th coordinate.
By this observation, we have that $\hat{\vex}$ must be feasible for $\II$ even if $\vex'$ was not, and $\tilde{\vex}$ must be feasible for $\II'$ even if $\vex$ was not.

% \paragraph{Case 2:} Otherwise, there are at least 2 elements with $i \in \suppo(\veg_j)$, meaning that $|x_i - x'_i| \geq 2$.
% Take $\bar{\veg} = \veg_j$ for any such $\veg_j$ with $i \in \suppo(\veg_j)$, and set $\hat{\vex} \df \vex' + \bar{\veg}$, $\veg \df \vex - \hat{\vex}$, and $\tilde{\vex} \df \vex' + \veg = \vex - \bar{\veg}$.
% Now let us distinguish two subcases: whether the new bound is tighter or looser than the original.

% \paragraph{Case 2a:} If $\vedelta = +\vece_i$, everything feasible for $\vel'$ is also feasible for $\vel$, thus straightforwardly $\vex'$ and $\hat{\vex}$ are feasible for $\vel$, but $\vex$ may not be feasible for $\vel'$, and we need to argue that $\vel' \leq \tilde{\vex}$.
% Since $\tilde{\vex} = \vex' + \veg$ and $g_i \geq 1$, $\tilde{\vex}$ is feasible for the modified instance, and we are done with the case $\vedelta = +\vece_i$.

% \paragraph{Case 2b:} Otherwise $\vedelta = -\vece_i$, thus $\vex$ and $\tilde{\vex}$ are feasible for the modified instance, but $\vex'$ may not be feasible for the original instance, and we need to argue that $\hat{\vex}$ is feasible for the original instance.
% Analogously, since $\hat{\vex} = \vex' + \bar{\veg}$ and $\bar{g}_i \geq 1$, $\hat{\vex}$ is feasible for the original instance.

% \medskip

The rest of the argument is identical to the one in the proof of Lemma~\ref{lem:prox-optima}.
We can write (in both cases)
$$\vex - \vex' = \veg + \bar{\veg} = (\tilde{\vex} - \vex') + (\hat{\vex} - \vex'),$$
which, using Proposition~\ref{prop:prox-superadd} with values $\vex = \vex'$, $\vey_1 = \veg$, $\vey_2 = \bar{\veg}$, gives
\begin{equation}
  f(\vex) - f(\tilde{\vex}) \geq f(\hat{\vex}) - f(\vex') \enspace .
\end{equation}
Since $\vex$ is an optimum of the original instance and $\tilde{\vex}$ is feasible for it, the left hand side is non-positive, and so is $f(\hat{\vex}) - f(\vex')$.
But since $\vex'$ is an optimum of the modified instance and $\hat{\vex}$ is feasible for it, $f(\hat{\vex}) - f(\vex')$ must be nonnegative, hence $f(\hat{\vex}) = f(\vex')$, thus $\hat{\vex}$ is another optimum of the modified instance.
However, by our assumption that there are at least two elements in the decomposition, both $\veg$ and $\bar{\veg}$ are non-zero, hence $\hat{\vex}$ is closer to $\vex$, a contradiction.
\qed
\end{proof}

%Let us give the main ideas behind the proof of Theorem~\ref{thm:basic-sensitivity};
%the full proof can be found in Appendix~\ref{app:sec:sensitivity}.
\sv{\paragraph{Proof Ideas.}
Consider the special case when~\(\vedelta\) is a unit vector, i.e.~\(\vedelta \in \{\vece_i, -\vece_i\}\), and let~\(\vex'\) be an optimum of the modified instance which is closest to~\(\vex\) in~\(\ell_1\)\=/norm.
As~\(\vex - \vex' \in \ker_\Z(A)\), it has a conformal decomposition~\(\vex - \vex' = \sum_{j} \veg_j\) with~\(\veg_j \in \G(A), \veg_j \sqsubseteq \vex - \vex'\) for every~\(j\) (see Proposition~\ref{prop:possum}).
We will show that in fact there can be only one element in the conformal decomposition.
If not, then we can take two distinct vectors of the conformal decomposition and use them to find another optimum solution~\(\hat{\vex}\) that is closer to~\(\vex\) than~\(\vex'\), and the approach would be similar as in the approach of the proof of Lemma~\ref{lem:prox-optima}.
Thus~\(\norm{\vex - \vex'}_1 \le g_1(A)\).
To prove it for any~\(\vedelta\), we can repeatedly apply the result where~\(\vedelta \in \{\vece_i, -\vece_i\}\) and use the triangle inequality, which gives the~\(\norm{\vedelta}_1\) factor in the bound in Theorem~\ref{thm:basic-sensitivity}.
}

\toappendix{
An analogous statement to Theorem~\ref{thm:basic-sensitivity} can be proved for the upper bound~\(\veu\):
\begin{corollary}
  \label{crly:basic-sensitivity-u}
  Let $\vex$ be an optimum of~\ref{IP}, $\vedelta \in \Z^n$, and $\veu' \df \veu + \vedelta$. Assuming that the IP with $\veu$ replaced by $\veu'$ is feasible, then it has an optimum $\vex'$ satisfying $\|\vex - \vex'\|_1 \leq \|\vedelta\|_1 g_1(A)$.
\end{corollary}
\begin{proof}
  Set $\bar{\vex} \df -\vex$, $\bar{\veu} \df -\vel$, $\bar{\vel} \df -\veu$, and~\(\bar{\vedelta} = -\vedelta\).
  Now apply Theorem~\ref{thm:basic-sensitivity} to $\bar{\vex}, \bar{\vel}, \bar{\veu}$ and $\bar{\vedelta}$.
  \qed
\end{proof}}

\medskip

Consider an instance of~\ref{IP} with $\veb = \vezero$.
For any $I \subseteq [n]$, the \emph{$I$-scaled IP} is an \ref{IP} instance with input $\bar{\vel}^I, \bar{\veu}^I, \bar{f}^I$ and with the constraint matrix $A^I$ defined as follows.
If $i \in I$, then $\bar{l}^I_i = \ceil{l_i / 2}$, $\bar{u}^I_i = \floor{u_i / 2}$ and $\bar{f}^I_i(x_i) = f_i(2 x_i)$; if $i \not\in I$, then $\bar{l}^I_i = l_i$, $\bar{u}^I_i = u_i$ and $\bar{f}^I_i(x_i) = f_i(x_i)$.
The $i$-th column of the constraint matrix $A^I \in \Z^{m \times n}$ is either twice the $i$-th column of $A$ if $i \in I$, otherwise it is simply the $i$-th column of $A$.
We use $I \triangle J$ to denote the symmetric difference of sets $I,J$, i.e., $I \triangle J = (I \setminus J) \cup (J \setminus I)$.
We denote by~\(\bar{A} \in \Z^{(n + m) \times 3n}\) the matrix~\(\left(\begin{smallmatrix}A & 2A & 0 \\ I & 2I & -I\end{smallmatrix}\right)\).
Our goal is to prove the following theorem:

\begin{theorem} \label{thm:sensitivity}
  Let $I, J \subseteq [n]$, and let $\vex^I$ be an optimum of the $I$-scaled IP.
  Then if the $J$-scaled IP is feasible, it has an optimum $\vex^J$ satisfying
  \begin{equation}
    \|\vex^I - \vex^J\|_1 \leq 2|I \triangle J| g_1(\bar{A}) \leq 4 |I \triangle J| g_1(A) \;.
  \end{equation}
\end{theorem}

The second inequality in Theorem~\ref{thm:sensitivity} follows from the following lemma.
\begin{lemma}
  \label{lem:graver-decomp}
  $g_1(\bar{A}) \leq 2g_1(A)$.
\end{lemma}
\begin{proof}
	%Let $\pi: \Z^n \to \Z^{3n}$ be the following mapping: for a vector $\veg \in \Z^n$ define $\pi(\veg) = (\vex, \vey, \vez) \in \Z^{3n}$ by setting, for each $i \in [n]$, $x_i = g_i \mod 2$, $y_i = \lfloor g_i / 2 \rceil$, and $z_i = g_i$.
  Consider an element of~\(\G(\bar{A}) \subset \Z^{3n}\) and split it evenly into three vectors~\(\vex, \vey, \vez \in \Z^n\) (i.e.~the first~\(n\)~components are~\(\vex\), the next~\(n\)~components are~\(\vey\), and the last~\(n\)~components are~\(\vez\)).
  Define~\(\veg \df \vex + 2\vey\).
  We have~\(\vex + 2\vey = \vez\) from the ``bottom half'' of~\(\bar{A}\) where~\(I\vex + 2I\vey = I\vez\).
  We claim that $\veg \in \G(A)$.
  For contradiction assume otherwise, meaning that~\(\veg\) can be decomposed as~$\veg = \veg' + \veg''$ with $\veg', \veg'' \in \Ker_\Z(A) \setminus \{\vezero\}$ and $\veg', \veg'' \sqsubseteq \veg$.
  We will define three new vectors~\(\vex', \vey', \vez' \in \Z^n\) and we will show that~\((\vex', \vey', \vez') \sqsubseteq (\vex, \vey, \vez)\), thus getting a contradiction with conformal minimality of an element of~\(\G(\bar{A})\).
  For~\(i \in [n]\) set~\(y'_i = \max\{0, \min\{y_i, \floor{g'_i/2}\}\}\) if~\(y_i\) is non-negative, and~\(y'_i = \min\{0, \max\{y_i, \ceil{g'_i/2}\}\}\) if~\(y_i\) is negative.
  It is clear that~\(\vey' \sqsubseteq \vey\).
  We set~\(\vex' = \veg' - 2\vey'\) and~\(\vez' = \veg'\).
  As~\(\veg'\) is non-zero, it follows that at least one of~\(\vex'\) and~\(\vey'\) are non-zero, and that~\(\vez'\) is non-zero.
  We get that~\(\vex' \sqsubseteq \vex\) and~\(\vez' \sqsubseteq \vez\) as~\(\veg' \sqsubseteq \vex + 2\vey\).
  Finally from~\(\vex + 2\vey = \vez\) we get~\((\vex', \vey', \vez') \sqsubseteq (\vex, \vey, \vez)\) which yields the desired contradiction.
  Thus we bound~\(\norm{(\vex, \vey)}_1 \le g_1(A)\) and~\(\norm{\vez}_1 \le g_1(A)\) which finishes the proof.
  \qed
  % Define $(\vex', \vey', \vez')$ by setting, for each $i \in [n]$, $y'_i = \min\{y_i, \lfloor g'_i / 2 \rceil\}$, $x'_i = g'_i - 2y'_i$, and $z'_i = g'_i$.
  % (We use the notation $\lfloor a \rceil$ for the operation which rounds down if $a$ is positive and rounds up if it is negative.)
  % As~\(g'_i\) is non-zero~\((\vex', \vey', \vez')\) is also non-zero.
  % From~\(\vex + 2\vey = \vez\) it follows that~$(\vex', \vey', \vez') \in \Ker_{\Z}(\bar{A})$, it is non-zero, and $(\vex', \vey', \vez') \sqsubseteq (\vex, \vey, \vez)$, contradicting the fact that $(\vex, \vey, \vez)$ is $\sqsubseteq$-minimal.

%	The other direction can be seen similarly: if $\veg \in \G(A)$, then any $(\vex, \vey, \veg)$ such that $\veg = \vex + 2\vey$ and $\vex, \vey$ have the same sign pattern is an element of $\G(\bar{A})$.
%	Otherwise, any conformal decomposition of $(\vex, \vey, \veg)$ would also give a decomposition of $\veg$, contradicting its $\sqsubseteq$-minimality.
\end{proof}

We want to use Theorem~\ref{thm:basic-sensitivity} to prove Theorem~\ref{thm:sensitivity}.
In order to do that, we will construct an auxiliary \ref{IP} instance which will be able to encode the $I$-scaled instance for any $I \subseteq [n]$ by only a small change in its bounds.

Specifically the constraint matrix is exactly~\(\bar{A}\).
%The solution vector $\vez$, the bounds $\hat{\vel}, \hat{\veu}$ and the objective function $\hat{f}$ can be divided into three blocks of dimension $n$, which we index by $\alpha$, $\beta$, and $\gamma$, e.g. $\vez = (\vez^{\alpha}, \vez^{\beta}, \vez^{\gamma})$.
% The bounds $\hat{\vel}^I, \hat{\veu}^I$ are defined as follows: for $i \in I$, $\hat{l}^I_i = \hat{u}^I_i = 0$, for $i \not\in I$, $\hat{l}^I_i = -1$ and $\hat{u}^I_i = 1$; in either case, $\hat{l}^I_{n+i} = \ceil{l_i / 2}$ and $\hat{u}^I_{n+i} = \floor{u_i / 2}$, and $\hat{l}^I_{2n +i} = l_i$ and $\hat{u}^I_{2n +i} = u_i$.
The bounds $\hat{\vel}^I, \hat{\veu}^I$ are defined as follows:
for~\(j \in [n]\), we set
\begin{itemize}
  \item $\hat{l}^I_{2n +j} = l_j$ and $\hat{u}^I_{2n +j} = u_j$,
  \item $\hat{l}^I_{n+j} = \ceil{l_j / 2}$ and $\hat{u}^I_{n+j} = \floor{u_j / 2}$,
  \item if~\(j \in I\), $\hat{l}^I_j = \hat{u}^I_j = 0$,
  \item otherwise~\(j \not\in I\), then~$\hat{l}^I_i = -1$ and $\hat{u}^I_i = 1$,
\end{itemize}
The objective function $\hat{f}$ is simply $\hat{f}(\vez) = f((z_i)_{i \in [2n+1, 3n]})$.
In summary:
\begin{align}
	\min & \left\{ \hat{f}( \vez^I ) : \ \bar{A} \vez^I = \vezero, \ \hat{\vel}^I \leq \vez \leq \hat{\veu}^I, \  \vez^I \in \Z^{3n} \right\}. \tag{aux-$I$-IP} \label{aux-I-IP}
\end{align}
Given a solution $\vex^I$ of the $I$-scaled IP, a vector $\vez^I \in \Z^{3n}$ intuitively encodes $\vex^I$ in the following way: if $i \not\in I$, the variable $i$ is already in the refined grid and we split its value into the least significant bit ($z^I_i$) and the rest ($z^I_{i+n}$); otherwise if $i \in I$, the variable $i$ is not in the refined grid, so we just set $z^I_{i+n} = x^I_{i}$.
Formally, if $i \in I$, set $z^I_{i+n} = x^I_i$ and $z^I_{i} = 0$, otherwise we need to distinguish whether $x^I_i$ is positive or negative: if $x^I_i \geq 0$, set $z^I_{i+n} = \floor{x^I_i / 2}$ and $z^I_i = x^I_i \mod 2$; otherwise set $z^I_{i+n} = \ceil{x^I_i / 2}$ and $z^I_i = -x^I_i \mod 2$.
For each $i \in [n]$, set $z^I_{i+2n} = z^I_{i} + 2 z^I_{n+i}$. 
In the other direction, given a solution $\vez^I$ of the~\ref{aux-I-IP}, we construct a solution $\vex^I$ of the $I$-scaled IP as follows.
If $i \in I$, set $x^I_i = z^I_{n+i}$, otherwise if $i \not\in I$, set $x^I_i = z^I_{i} + 2z^I_{n+i}$.
Notice that the value of $\vez^I$ with respect to $\hat{f}$ is equal to the value of $\vex^I$ with respect to $\bar{f}^I$:
\begin{itemize}
  \item \(\forall i \in I: \, \bar{f}^I_i(x^I_i) = f_i(2x^I_i) = \hat{f}_i(z^I_{2n+i})\) since~\(z^I_{i+n} = x_i^I,\, z_i^I = 0\) and~\(z^I_{i+2n} = z^I_i + 2z^I_{n+i} = 2x^I_i\),
  \item \(\forall i \not\in I: \, \bar{f}^I_i(x^I_i) = f_i(x^I_i) = \hat{f}_i(z^I_{2n+i})\) since~\(z^I_{i+2n} = z^I_i + 2z^I_{n+i} = x^I_i\) by definition.
\end{itemize}
By the construction above we immediately get:
\begin{lemma}
  \label{lem:aux-I-IP_preserves_optima}
  A solution $\vex^I$ of the $I$-scaled IP is optimal if and only if the corresponding solution $\vez^I$ of~\ref{aux-I-IP} is optimal.
\end{lemma}
%Now we are ready to prove Theorem~\ref{thm:sensitivity}:
\begin{proof}[of Theorem~\ref{thm:sensitivity}]
Let $\vez^I$ be the solution of~\ref{aux-I-IP} corresponding to the optimum~$\vex^I$ from the statement.
By Lemma~\ref{lem:aux-I-IP_preserves_optima}, \(\vez^I\) is optimal for~\ref{aux-I-IP}.
Observe that when we replace $I$ with $J$, the bounds of the~\ref{aux-I-IP} instance only change in coordinates $I \triangle J$, and in each coordinate of~\(I \triangle J\) it changes by $2$ (for a coordinate in $J \setminus I$, the lower bound changes from $0$ to $-1$, and the upper bound from $0$ to $+1$, and for a coordinate in $I \setminus J$, the bounds change from $+1$ or $-1$ to $0$).
Thus, by Theorem~\ref{thm:basic-sensitivity}, there exists an optimum $\vez^J$ at $\ell_1$-distance at most $2|I \triangle J| g_1(\bar{A})$.
Finally, the corresponding optimum $\vex^J$ of the $J$-scaled IP is at the claimed distance, again, by construction.
\qed
\end{proof}

%Finally, we will need the next lemma in the algorithmic part of Theorem~\ref{thm:dual-alg}.
% As the proof is quite simple, we defer it to Appendix~\ref{app:sec:sensitivity}.

%Sensitivity in integer and linear programming is the notion of how much must an optimal solution of an~\ref{IP} change when some part of the input changes, such as when $\veb, \vel, \veu$ or the objective function changes.
%It is known that if $\vex$ is an optimum of the original~\ref{IP}, and $\veb$ changes by $\delta$, there exists an optimal solution $\vex'$ of the new IP within $\ell_p$-distance $\delta g_p(A)$ of $\vex$, for any $p$ with $1 \leq p \leq +\infty$.
%%It is also not too hard to show a similar claim for changes in the box constraints $\vel, \veu$.

% For the purposes of a fast dual algorithm (Section~\ref{sec:dual-alg}) we are interested in a somewhat different type of change.
% Instead of going from the lattice $2s \Z^n$ to the lattice $s \Z^n$ in one ``jump'', we can refine one variable at a time, i.e. we can change the bounds on $x_i$ from $\ceil{l_i / 2s}$ and $\floor{u_i / 2s}$ to $\ceil{l_i / s}$ and $\floor{u_i / s}$ one by one.
% The question we would like to answer is, given an optimal solution $\vex$, after refining one variable, how close is some optimal solution $\vex'$ of this new instance?

By Lemma~\ref{lem:graver-decomp}, we can use asymptotically the same bounds for $g_1(\bar{A})$ as $g_1(A)$; for example, we have $g_1(\bar{A}) \leq 2g_1(A) \leq (2\|A\|_\infty)^{2^{\Oh(\td_D(A))}}$~\cite[Theorem 3]{MOR}.
The next lemma analogously relates $\td_D(A)$ and $\td_D(\bar{A})$.

\begin{lemma} \label{lem:td-decomp}
	Let $F$ be a $\td$-decomposition of $G_D(A)$. %and let\\ $K=\max_{P \text{: root-leaf path in }F}\prod_{i=1}^{\ttd(F)} (k_i(P)+1)$.
	A $\td$-decomposition $\bar{F}$ of $G_D(\bar{A})$ with $\height(\bar{F}) = \height(F) + 1$ and $\ttd(\bar{F}) = \ttd(F) + 1$ can be computed in linear time.
\end{lemma}
\begin{proof}
    Start from $F$, and go over $i=1, \dots n$.
    Let $M$ be the set of rows of $A$ which are non-zero in column $i$.
    Observe that all of them must lie on some root-leaf path in $F$, and call $v_i$ the leaf of this path.
    Attach a new leaf to $v_i$, representing the row $m+i$ of $\bar{A}$.
    It is easy to verify that $\bar{F}$ is a $\td$-decomposition of $\bar{A}$.

    Since we have only been attaching leaves to vertices which were leaves in $F$, we have only possibly increased the height by $1$, and we have only possibly added one more non-degenerate node on any root-leaf path, increasing the topological height by at most $1$.
    This shows the lemma.
    \qed
\end{proof}

%% file: dual-algorithm.tex
% TEX root = main.tex
\section{Dual algorithm} \label{sec:dual-alg}
\sv{\toappendix{\section{Additional Material for Section~\ref{sec:dual-alg}} \label{app:sec:dual-alg}}}

%In this section, we will show a near-linear algorithm for the case of bounded dual treedepth.
The main technical ingredient of our algorithm is a dynamic data structure solving the following subproblem: for an~\ref{IP} instance $(A, f, \vezero, \vel, \veu)$, its feasible solution $\vex \in \Z^n$, and a parameter $\rho \in \N$, the \ref{l1-IP}~problem is to solve
\begin{equation}
	\min \{f(\vex+\veh) \mid A\veh = \vezero, \, \vel \leq \vex+\veh \leq \veu, \, \|\veh\|_1 \leq \rho, \, \veh \in \Z^n \} \enspace . \tag{$\ell_1$-IP} \label{l1-IP}
\end{equation}
If two~\ref{l1-IP} instances differ in at most $\sigma$ coordinates of their bounds and objective functions, we call them \emph{$\sigma$-similar}.
%Two instances $(\lambda_1, \vex_1)$ and $(\lambda_2, \vex_2)$ of~\eqref{AugIP} are defined as \emph{$\sigma$-similar} if $\lambda_1 = \lambda_2$ and $\mathopen|\suppo(\vex_1 - \vex_2)| \leq \sigma$.
We shall construct a data structure we call ``convolution tree'' which maintains a representation of an~\ref{l1-IP} instance together with its optimal solution, takes linear time to initialize, and takes time roughly~$\bigO(\sigma \log n)$ to update to represent a $\sigma$-similar~\ref{l1-IP} instance.

Before we give a definition of a convolution tree, let us first explain the problem it solves.
The~\ref{l1-IP} problem is implicitly present in the literature as an important subtask of algorithms for~\ref{IP} with small $\td_D(A)$~\cite{HemmeckeOR13,MOR}.
A feasible solution~\(\veh\) to an instance of~\ref{l1-IP} should be viewed as a linear combination of columns of~\(A\) with coefficients given by~\(\veh\) such that the linear combination of the columns is~\(\vezero\), and the cost of the solution is~\(f(\vex + \veh)\).
Finding an optimum solution can be done by the following dynamic programming approach:
For simplification let us assume~\(\vex = \vezero\) and that~\(\vel, \veu\) are the all (minus) infinity vectors.
For every~\(1 \le i \le j \le n\) we denote by~\({f\mid}_{i, j}\) the restriction of~\(f\) to coordinates in~\([i, j]\), and~\({f\mid}_j \coloneq {f\mid}_{1, j}\).
For any linear combination~\(\sum_{i = 1}^j c_i A_{\bullet, i}\) of any first~\(j\)~columns of~\(A\) where~\(c_i \in [-\rho, \rho]\) for each~\(i \in [j]\) we have~\(\|\sum_{i = 1}^j c_i A_{\bullet, i}\|_\infty \le \rho \|A\|_\infty\).
The restriction on the domain of the coefficients~\(c_1, \ldots, c_j\) comes from the condition~\(\|\veh\|_1 \le \rho\) in~\ref{l1-IP}.
Thus for every~\(j \in [n]\), and vector~\(\ves \in [-\rho\|A\|_\infty, \rho\|A\|_\infty]^m\), we can compute the cheapest linear combination of the first~\(j\)~columns that sums up to~\(\ves\) recursively -- for every~\(c \in [-\rho, \rho]\), compute the cheapest linear combination~\(c_1^c, \ldots, c_{j - 1}^c\) of the first~\(j - 1\)~columns of~\(A\) that sums up to~\(\ves - cA_{\bullet, j}\), and store the~\(c\) that minimizes~\({f\mid}_j((c_1^c, \ldots, c_{j - 1}^c, c))\).
To bound the running time of this dynamic programming algorithm, observe that the dynamic programming table has~\((2\rho\|A\|_\infty + 1)^m n\) entries and computing each of them requires time~\(\bigO(2\rho\|A\|_\infty + 1)\).

As we have sketched in the beginning of Section~\ref{sec:sensitivity}, the algorithm for dual treedepth will not descale all variables at once but do it one by one instead.
This is where Theorem~\ref{thm:sensitivity} comes into play:
if we write the initial~\ref{IP} instance as an~\ref{aux-I-IP} instance with optimum~\(\vex^*\) (which can be computed by the dynamic programming approach described above), then the optimum of an instance where we descale one variable has the property that there exists an optimum~\(\vex'\) with~\(\|\vex^* - \vex'\| \le 4g_1(A)\).
Thus computing~\(\vex'\) can be viewed as solving~\ref{l1-IP} with~\(\vex = \vex^*\) and~\(\rho = 4g_1(A)\).

We will need to solve~\(n - 1\)~\ref{l1-IP}~instances in order to solve a single~\ref{IP}~instance of the~\(\lceil\log\|\veu - \vel\|_\infty\rceil + 1\)~many~\ref{IP}~instances of the scaling algorithm.
If we solve each of them using the dynamic programming approach, then in the end, the dependence on~\(n\) in the running time is~\(\bigO(n^2)\).

Instead of solving each~\ref{l1-IP}~instance from scratch, we shall leverage the fact that subsequent~\ref{l1-IP}~instances in a single iteration of the scaling algorithm do not differ too much which leads to the possibility of \emph{sparse updates}.
To simplify the subsequent explanation, let us now assume that the update operation only affects the functions~\(f_i\) as we can simulate the updates to the bounds~\(l_i\) and~\(u_i\) by changes to~\(f_i\): when~\(u_i\) is updated to~\(K\), we can set~\(f_i(a) \approx \infty\) for every~\(a \ge K\);
we need to be careful to keep~\(f_i\) convex though so instead we set~\(f_i(a) = f_i(K) + M(a - K)\) for some sufficiently large~\(M\) so that for every~\(\vex \not\in [\vel, \veu]\) and~\(\vex' \in [\vel, \veu]\) we have~\(f(\vex) > f(\vex')\).
We can handle updates of~\(l_i\) analogously.
A change in the objective function~\(f_i\) means that we change the cost of all partial solutions on columns~\([i, n]\).
A naive way to perform an update operation would be to recompute the dynamic programming table but this would take time~\(\Omega(n)\).
The idea behind a fast update is to build an analogue of a segment tree~\cite{cg-book} on top of the dynamic programming table.

For simplicity assume that~\(n\) is a power of two, otherwise we can add enough all\=/zero columns to make it so.
We build a rooted full binary tree~\(\mathcal{T}\) with~\(n\)~leaves, meaning that every internal vertex has exactly two children and we denote the root by~\(r\).
Note that the height of~\(\mathcal T\) is~\(1 + \log n\).
Notationally, let~\(r\) be at depth~0 and the leaves at depth~\(\log n\).
We enumerate leaves of~\(\mathcal T\) with numbers~\([n]\) from left to right.

For each vertex~\(u \in V(\mathcal{T})\) we say that it \emph{represents} an interval~\([i, j]\) if the leaves in the subtree of~\(\mathcal T\) rooted at~\(u\) have labels exactly~\([i, j]\).
For each vertex~\(u \in V(\mathcal{T})\) and each~\(\ves \in [-\rho\|A\|_\infty, \rho\|A\|_\infty]^m\) we shall compute the cheapest linear combinations of columns~\(A_{\bullet, i}, \ldots, A_{\bullet, j}\) with respect to~\({f\mid}_{i, j}\) that sums up to~\(\ves\) or let the cost be~\(\infty\) if no such linear combination exists, and we denote this value by~\(\mathcal{T}(u, \vev)\).
The computation shall be done as follows.
Let us assume that for every vertex~\(u \in V(\mathcal{T})\) and each~\(\ves \in [-\rho\|A\|_\infty, \rho\|A\|_\infty]^m\) we have implicitly~\(\mathcal{T}(u, \ves) = \infty\).
For every leaf~\(u\) representing the interval~\([i, i]\) of~\(\mathcal{T}\) and every~\(c \in [-\rho\|A\|_\infty, \rho\|A\|_\infty]\) we set~\(\mathcal{T}(u, cA_{\bullet, i}) = f_i(c)\).
Now successively for depths~\(\log n - 1, \log n - 2, \ldots, 0\) and for every vertex~\(u\) at these depths with children~\(u_1\) and~\(u_2\), and for every~\(\ves_1, \ves_2 \in [-\rho\|A\|_\infty, \rho\|A\|_\infty]^m\) we set
\begin{equation}
  \mathcal{T}(u, \ves_1 + \ves_2) = \min(\mathcal{T}(u, \ves_1 + \ves_2), \mathcal{T}(u_1, \ves_1) + \mathcal{T}(u_2, \ves_2))\; .
\end{equation}

The reason why we call our data structure convolution tree is that we can also view computing~\(\mathcal{T}(u, \ves)\) as a~\((\min, +)\)\=/convolution
\begin{equation}
  \min_{\ves_1, \ves_2;\; \ves_1 + \ves_2 = \vev} \mathcal{T}(u_1, \ves_1) + \mathcal{T}(u_2, \ves_2)\; .
\end{equation}
We can also store the linear combination of columns whose indices are represented by a vertex~\(u \in V(\mathcal{T})\) that make up~\(\mathcal{T}(u, \ves)\) for every~\(\ves \in [-\rho\|A\|_\infty, \rho\|A\|_\infty]^m\).
To retrieve the solution of~\ref{l1-IP} we need to simply query~\(\mathcal{T}(r, \vezero)\).

When a function~\(f_i\) is updated, then it is only necessary to recompute~\(\mathcal{T}(u, \cdot)\) for vertices~\(u\) that lie on the path from the leaf of~\(\mathcal{T}\) labelled by~\(i\) to the root.
The recomputation when~\(f_i\) is updated is done exactly as the initialization -- by computing a convolution in every vertex on a root\=/leaf path from the leaf labelled~\(i\).
This update only affects~\(\log n + 1\) vertices and each convolution requires time~\([-\rho\|A\|_\infty, \rho\|A\|_\infty]^{2m}\), and the initialization takes time~\(\bigO(n)\) (ignoring dependence on parameters and~\(m\)).

When we formalize the convolution tree, we shall show how to decrease the super\=/polynomial dependence on~\(m\) to~\(\td_D\).
The idea here is that we can leverage the sparse recursive structure of the~\(\td\)\=/decomposition~\(F\) of~\(A\) instead of considering each column separately.
Essentially, notice that for every~\(\veh\) such that~\(A\veh = \vezero\) and any~\(i \in [n]\), the prefix sum~\(\sum_{j = 1}^i A_{\bullet, i}h_i\) is non\=/zero in at most~\(\td_D(A)\) coordinates, and thus can be represented very compactly.
This eventually decreases the exponent to~\(2\height(F) + 1\).

\begin{definition}[Convolution Tree]
  \label{def:convol_tree}
  Let $A \in \Z^{m \times n}$, $F$ be a $\td$-decomposition of $G_D(A)$, $\rho \in \N$, $R \df [-\rho \|A\|_\infty, \rho \|A\|_\infty]^{k_1(F)}$ and $R' \df R \times [0,\rho]$.
  A \emph{convolution tree} is a data structure $\T$ which stores two vectors $\vel_{\T}, \veu_{\T} \in \Z^n$ and a separable convex function $f_{\T}\colon \R^n \to \R$, and we call $\vel_{\T}, \veu_{\T}, f_{\T}$ the \emph{state of $\T$}.
  A convolution tree  $\T$ supports the following operations:
  \begin{enumerate}
    \item
      \textsc{Init}$(\vel_{\T}, \veu_{\T}, f_{\T})$ initializes $\T$ to be in state $\vel_{\T}, \veu_{\T}, f_{\T}$.
    \item
      \textsc{Update}$(i,l_i,u_i,f_i)$ is defined for $i \in [n]$, $l_i,u_i \in \Z$ and a univariate convex function $f_i: \R \to \R$.
      Calling \textsc{Update}$(i, l_i, u_i, f_i)$ changes the $i$-th coordinates of the state into $l_i, u_i$ and $f_i$, i.e.,
      if $\vel_{\T}, \veu_{\T}, f_{\T}$ is the state of $\T$ before calling \textsc{Update}$(i, l_i, u_i, f_i)$ and $\vel'_{\T}, \veu'_{\T}, f'_{\T}$ is the state of $\T$ afterwards, then for all $j \in [n] \setminus \{i\}$, $(\vel'_{\T})_j=(\vel_{\T})_j$, $(\veu'_{\T})_j=(\veu_{\T})_j$, and $(f'_{\T})_j=(f_{\T})_j$, and $(\vel'_{\T})_i = l_i$, $(\veu'_{\T})_i = u_i$, and $(f'_{\T})_i = f_i$.
      %\byasaf{What does updating an oracle mean?}
    \item
      \label{it:conv_tree_veG}
      \textsc{Query} returns a solution sequence~$\veG \in (\Z^{n} \cup \{\undffd\})^{R'}$ where~${\veg_{\ver,\eta} \coloneqq (\veG)_{\ver,\eta}}$ is a solution of
      \begin{equation}
        \min \left\{f(\veg) \mid A\veg = (\ver, \vezero), \, \vel_{\T} \leq \veg \leq \veu_{\T}, \, \veg \in \Z^n, \|\veg\|_1 = \eta\right\}, \label{eq:convol_tree_subproblem}
      \end{equation}
      with~\(\vezero\) the \((m - k_1(F))\)\hy{}dimensional zero vector, and~\(\vel_{\T}, \veu_{\T}, f_{\T}\) the current state of~\(\T\).
      If~\eqref{eq:convol_tree_subproblem} has no solution, we define its solution to be $\undffd$.
      Each vector~\(\veg_{\ver, \eta}\) has a sparse representation as there are at most~\(\eta\)~coordinates that are non-zero due to the constraint~\(\norm{\veg}_1 = \eta\), and thus can be represented by a list of length~\(\eta\) of index-value pairs.
  \end{enumerate}
\end{definition}
In Definition~\ref{def:convol_tree}, the set $R$ represents the set of possible right hand sides of~\eqref{eq:convol_tree_subproblem}, and the extra coordinate in the set $R'$ serves to represent the set of considered $\ell_1$-norms, which includes~0, of solutions of~\eqref{eq:convol_tree_subproblem}.
We are being slightly opaque in the definition of \textsc{Update} since we are only given a comparison oracle for~\(f(\vex) = \sum_{i = 1}^n f_i(x_i)\) and not individual~\(f_i\)'s.
However, a comparison oracle for~\(f_i\)'s is implicit from the comparison oracle for~\(f_i\) thus we can speak about updating individual~\(f_i\)'s.
We also introduce the following shortcut: for~\(\sigma \in [n]\) by~\emph{\(\sigma\)\=/\textsc{Update\((U, \vel_U, \veu_U, f_U)\)}} where~\(U \subseteq [n]\) with~\(|U| = \sigma\), \(\vel_U, \veu_U \in \Z^U\), and~\(f_U\colon \R^U \to \R\) we mean calling~\(\textsc{Update\(((U)_i, (\vel_U)_i, (\veu_U)_i, (f_U)_i)\)}\) for each~\(i \in [\sigma]\) in increasing order of~\(i\).

We present an implementation of the convolution tree in the following lemma.
We prove in Appendix~\ref{sec:convolution-tree-opt} that the dependence on~\(n\) in the running time of each operation is essentially tight in the comparison model.

\begin{lemma}[Convolution Tree Lemma] \label{lem:convol_tree}
  Let $A,F,\rho$, and $\T$ be as in Definition~\ref{def:convol_tree}, and~\(H = 2\height(F) + 1\).
  We can implement all operations so that the following holds.
	\begin{enumerate}
		\item \textsc{Init}$(\vel_{\T}, \veu_{\T}, f_{\T})$ can be realized in time $\bigO\left((2\|A\|_\infty \cdot \rho + 1)^H \cdot 2n\right)$,
		\item \textsc{Update}$(i,l_i,u_i,f_i)$ can be realized in time~\(\bigO\left((2\|A\|_\infty \cdot \rho + 1)^H\ttd(F) \log n\right)\),
		\item \textsc{Query} can be realized in time~\(\Oh(|R|\rho^2)\) and the result can be represented in space~\(\Oh(|R|\rho^2)\).
                  For any~\((\ver, \eta) \in R'\) the vector~\(\veg_{\ver, \eta}\) can be retrieved in time~\(\Oh(\rho)\) and represented in space~\(\Oh(\rho)\).
	\end{enumerate}
\end{lemma}
Note that the running time of \textsc{Query} does not depend on~\(n\).
% For notational convenience for~\(\sigma \in [n]\) ~\(\sigma\)\=/\textsc{Update\((U, \vel_U, \veu_U, f_U)\)} defined for~\(U \subseteq [n]\) with\(|U| = \sigma\) \sigma \in [n]\) where~\(U \subseteq [n]\), \(|U| = \sigma\), \(\vel_U, \veu_U \in \Z^U\), and a s
Given Lemma~\ref{lem:convol_tree} whose proof we postpone, we prove Theorem~\ref{thm:dual-alg} as follows.
\begin{proof}[of Theorem~\ref{thm:dual-alg}]
	The algorithm consists of two nested loops.
        The outer loop is essentially the same as the one of the scaling algorithm: we begin by setting $k = \ceil{\log_2 \norm{\vel, \veu}_\infty} + 1$, recentering the instance using an initial feasible solution, and setting $\vex_k = \vezero$ as the optimum of the $2^k$-scaled instance.
        If an initial feasible solution is not given as part of the input, it can be found using Lemma~\ref{lem:initial-feasible-sol}.
        Similarly to the case of Theorem~\ref{thm:primal-alg}, finding the initial feasible solution is the only reason why the running time has the~\(\log\norm{\veb}_\infty\) factor.

	We call each iteration of the outer loop a \emph{phase}, and in each phase $k$ is decremented, a solution $\vex_k$ of the $2^k$-scaled instance is obtained, until $k=0$ and $\vex_k$ is the optimum of the original instance.
	However, the process by which the algorithm goes from $\vex_k$ to $\vex_{k-1}$ is different.
	
	In each phase, we begin by constructing the~\ref{aux-I-IP} instance for $I=[n]$, centered at the feasible solution $\vez_{k+1}$ derived from $\vex_{k+1}$ from the previous phase.
	Each phase has $n$ \emph{steps}.
	In step $j$, we begin with a solution which is optimal for the $[j,n]$-scaled instance, that is, we have already refined coordinates $1$ through $j-1$, and the task of step $j$ is to refine the $j$-th coordinate, that is, find the optimum of the $[j+1,n]$-scaled instance.
	By Theorem~\ref{thm:sensitivity}, this optimum cannot differ from the previous one by more than $2g_1(\bar{A})$ in the $\ell_1$-norm.
	Call $\vez^{j-1}$ the optimum from the previous step, and $\vez^{j}$ the nearby optimum to be found.
	The proximity bound implies that there is a step $\veh = \vez^{j} - \vez^{j-1}$ of small $\ell_1$-norm.
	We recenter the~\ref{aux-I-IP} for $I = [j+1,n]$ at $\vez^{j-1}$, which means that we are looking for an optimum of~\ref{aux-I-IP} with an additional $\ell_1$-norm bound $2g_1(\bar{A})$.
        We can do this quickly using the convolution tree as follows.
	
	At the beginning of each phase, let $I = [n]$, and initialize a convolution tree $\T$ with the matrix $\bar{A}$, a $\td$-decomposition $\bar{F}$ from Lemma~\ref{lem:td-decomp}, bounds $\hat{\vel}^I, \hat{\veu}^I$, right hand side $\vezero$, objective function $\hat{f}$, and $\rho = 2 g_1(\bar{A})$.
	Then, perform the following for $j=1, \dots, n$ in increasing order of~\(j\):
	Let $\veh^j$ be the solution of the~\ref{l1-IP} represented by $\T$, $\sigma = |\suppo(\veh^j)|$, let $\vez^j \df \vez^{j-1} + \veh^j$, and call a $\sigma$-\textsc{Update} on $\T$ to recenter its~\ref{l1-IP} at $\vez^j$.
	Now call an \textsc{Update} on the coordinate $j+1$ to change its bounds from $0,0$ to $-1,+1$.
	
        Notice that at the beginning of step $j$, $\T$ represents the~\ref{aux-I-IP} for $I = [j,n]$, and since it is centered at the optimum of the~\ref{aux-I-IP} for $I = [j-1,n]$ and by Theorem~\ref{thm:sensitivity} and our choice of $\rho$, the optimum obtained by calling~\textsc{Query} on~\(\T\) is in fact the optimum of the~\ref{aux-I-IP} for $I = [j,n]$.
	Thus, after the $n$-th step, the optimum of $\T$ is the optimum of the~\ref{aux-I-IP} for $I = \emptyset$, and we construct from it the solution $\vex_{k-1}$ used to initialize the next phase.
	This, together with the correctness of the scaling algorithm, shows the correctness of the algorithm.
	
        As for complexity, notice that there are $\Oh(\log \|\veu-\vel\|_\infty)$ phases, each phase has $n$ steps, each step consists of calling a $\sigma$-\textsc{Update} followed by a~\textsc{Query} with $\sigma \leq g_1(\bar{A})$, and one additional~\textsc{Update} followed by a~\textsc{Query}.
	Plugging into Lemma~\ref{lem:convol_tree} finishes the proof.

        Note that similarly to the primal algorithm (Theorem~\ref{thm:primal-alg}), the dependence on~\(\log\norm{\veb}_\infty\) is only necessary to find an initial feasible solution.
        \qed
\end{proof}

Before we prove Lemma~\ref{lem:convol_tree}, we formalize the notion of convolution we use.
\begin{definition}[Convolution]
  Given a set $R \subseteq \Z^\delta$ and tuples $\vealpha=\left(\alpha_{\ver}\right)_{\ver \in R}, \vebeta = \left(\beta_{\ver}\right)_{\ver \in R} \in \left(\Z \cup \{+\infty\}\right)^R$, a tuple $\vegamma = \left(\gamma_{\ver}\right)_{\ver \in R} \in \left(\Z \cup \{+\infty\}\right)^R$ is the \emph{convolution of $\vealpha$ and $\vebeta$}, denoted $\vegamma = \convol(\vealpha, \vebeta)$, if
  \begin{equation}
    \gamma_{\ver} = \min_{\substack{\ver', \ver'' \in R\\\ver' + \ver'' = \ver}} \alpha_{\ver'} + \beta_{\ver''} \qquad \forall \ver \in R \enspace .
  \end{equation}
  A tuple of pairs $w(\vegamma)_{\vealpha, \vebeta} \in (R\times R)^R$ is called a \emph{witness of $\vegamma$ w.r.t. $\vealpha, \vebeta$} if
  \begin{equation}
    w(\vegamma)_{\vealpha, \vebeta}(\ver) = (\ver',\ver'') \Leftrightarrow \gamma_{\ver} = \alpha_{\ver'} + \beta_{\ver''} \qquad \forall \ver \in R \enspace .
  \end{equation}
\end{definition}
%It is straightforward (as noted, e.g., by Jansen and Rohwedder~\cite{JansenR:2018}) to index the sequences by some set $R \subseteq \Z^\delta$  instead of $[0,n-1]$.

\begin{proof}[of Lemma~\ref{lem:convol_tree}]
	We define $\T$ recursively over $\ttd(F)$, then describe how the operations are realized, and finally analyze the time complexity.
        If $\ttd(F) \geq 2$, we assume $A$ is dual block-structured along $F$ (recall Definition~\ref{def:dual-decomp}) and we have, for every $i \in [d]$, matrices $A_i, \bar{A}_i, \hat{A}_i = \binom{\bar{A}_i}{A_i}$ and a $\td$-decomposition $\hat{F}_i$ of $G_D(\hat{A}_i)$ with $\ttd(\hat{F}_i) < \ttd(F)$ and $\height(\hat{F}_i) \leq \height(F)$, and a corresponding partitioning of $\vel, \veu, \veg$ and $f$.
	If $\ttd(F) = 1$, let $d \df n$, $\bar{A}_i \df A_{\bullet, i}$ for each $i \in [d]$, and $A_1, \dots, A_d$ be empty.
	For $i,j \in [d]$, $i \leq j$, denote by $A[i,j]$ the submatrix of $A$ induced by the rows and columns of the blocks $\bar{A}_i, \dots, \bar{A}_j$, $\vel_{\T}[i,j] \df (\vel^i_{\T}, \dots, \vel^j_{\T})$, $\veu_{\T}[i,j] \df (\veu^i_{\T}, \dots, \veu^j_{\T})$, and $f_{\T}[i,j]$ be the restriction of $f_{\T}$ to the coordinates of $\vel_{\T}[i,j]$.
	Observe that $\hat{A}_i = A[i,i] = \left(\begin{smallmatrix}\bar{A}_i \\ A_i\end{smallmatrix}\right)$.
	
	We obtain $\T$ by first defining a convolution tree $\T_i$ for each $i \in [d]$, where $\T_i$ is a convolution tree for the matrix $\hat{A}_i$ and for $\vel_{\T}^i, \veu_{\T}^i$ and $f_{\T}^i$, and then constructing a binary tree $T$ whose leaves are the $\T_i$'s and which is used to join the results of the $\T_i$'s in order to realize the operations of $\T$.
	Since, for each $i \in [d]$, $\T_i$ is supposed to be a convolution tree for a matrix $\hat{A}_i$ with $\ttd(\hat{F}_i) < \ttd(F)$, if $\ttd(F) \geq 2$, we will construct $\T_i$ by a recursive application of the procedure which will be described in the text starting from the next paragraph.
	Now we describe how to obtain $\T_i$ if $\ttd(F)=1$, i.e., when $\hat{A}_i$ is just one column.
	When $\T_i$ is initialized or updated, we construct the sequence $\veG_i$ by using the following procedure.
	For each $(\ver, \eta) \in R'$, defining $\veg^i$ to be such $\veg^i \in \left(\{-\eta, \eta\} \cap [\vel_{\T}^i, \veu_{\T}^i]\right)$ which satisfies $\hat{A}_i \veg^i = (\ver, \vezero)$ and minimizes $f_{\T}^i(\veg^i)$, and returning either $\veg^i$ if it is defined or $\undffd$ if no such $\veg^i$ exists.
	Notice that $\veg^i, \vel^i_{\T}$ and $\veu^i_{\T}$ are scalars.
	
      We say that a rooted binary tree is \emph{full} if each vertex has $0$ or $2$ children, and that it is \emph{balanced} if its height is at most $\log(h) + 3$ where~\(h\) is the number of leaves of the tree.
	It is easy to see that for any number $h$ there exists a rooted balanced full binary tree with $h$ leaves.
	Let $T$ be a rooted balanced full binary tree with $d$ leaves labeled by the singletons $\{1\}, \dots, \{d\}$ in this order, and whose internal vertices are labeled as follows: if $u \in T$ has children $v,w$, then $u=v \cup w$; hence the root $r$ satisfies $r=[d]$ and the labels are subsets of consecutive indices in~$1, \dots, d$.
	We obtain $\T$ by identifying the leaves of $T$ with the roots of the trees $\T_i$ and will explain how to use $T$ to join the results of all $\T_i$'s in order to realize the operations of $\T$.

	To initialize $\T$ with vectors $\vel_\T$ and $\veu_\T$ and a function $f_{\T}$, we shall compute, in a bottom-up fashion, the sequence $\veG$ described in point~\ref{it:conv_tree_veG} of Definition~\ref{def:convol_tree}.
	Let $u$ be a node of $T$ and let $i \df \min u$ and $j \df \max u$ be the leftmost and rightmost leaves of $T_u$ (i.e., the subtree of $T$ rooted at $u$), respectively.
	Moreover, if $u$ is an internal node, let its left and right child be $v$ and $w$, respectively, and let $k \df \max v$ be the rightmost leaf of $T_v$.
	Consider the following auxiliary problem, which is intuitively problem~\eqref{eq:convol_tree_subproblem} restricted to blocks $i$ to $j$: simply append ``$[i,j]$'' to all relevant objects, namely $f_{\T}, \veg, A, \vel_{\T}$ and $\veu_{\T}$:
	\begin{multline}
		\min \big\{f_{\T}[i,j](\veg[i,j]) \mid A[i,j] \veg[i,j] = (\ver, \vezero), \, \vel_{\T}[i,j] \leq \veg[i,j] \leq \veu_{\T}[i,j], \\
		\veg[i,j] \in \Z^{n_i + \cdots + n_j}, \, \|\veg[i,j]\|_1 = \eta \big\}, \label{eq:conv_tree_aux}
	\end{multline}
	where $\vezero$ has dimension $\sum_{\ell=i}^j m_{\ell}$.
	Let $\veg^u_{\ver, \eta}$ be a solution of~\eqref{eq:conv_tree_aux} and let $\veG^u$ be the sequence $\left(\veg^u_{\ver'}\right)_{\ver' \in R'}$.
	
	If $u$ is a leaf of $T$, the sequence $\veG^u$ is obtained by querying $\T_u$ (which was defined previously).
	Otherwise, compute the sequences $\veG^v$ and $\veG^w$ for the children $v,w$ of $u$, respectively.
	Then, we compute a convolution $\vegamma^u$ of sequences, $\zeta \in \{v,w\}$, $\vegamma^\zeta$ obtained from $\veG^\zeta$ by setting $\gamma^\zeta_{\ver'} \df f_{\T}[\min \zeta, \max \zeta](\veG^\zeta_{\ver'})$, where $f_{\T}[\min \zeta, \max \zeta](\undffd) \df +\infty$.
	We also compute a witness $w(\vegamma^u)_{\vegamma^v, \vegamma^w}$ of $\vegamma^u$.
	The desired sequence $\veG^u$ is easily obtained from $w(\vegamma^u)_{\vegamma^v, \vegamma^w}$.
	
	The \textsc{Update} operation is realized as follows.
	Let $i, l_i, u_i, f_i$ be as described in Definition~\ref{def:convol_tree}.
	Let $\iota(i) \in [d]$ be the index of the block containing coordinate $i$, and let $i' \in [n_i]$ be the coordinate of block $\iota(i)$ corresponding to $i$.
	First traverse $T$ downward from the root to leaf $\iota(i)$, call \textsc{Update}$(i, l_i, u_i, f_i)$ on subtree $\T_{\iota(i)}$, and then recompute convolutions $\vegamma^w$ and sequences $\veG^w$ for each vertex $w$ on a root-leaf path starting from the leaf~\(\iota(i)\) and going towards the root.
	% The $\sigma$-\textsc{Update} operation is realized by simply calling \textsc{Update} for each $i \in U$, in increasing order of $i$.
	
	Let us analyze the time complexity.
	The convolution tree $\T$ is composed of $\ttd(F)$ levels of smaller convolution trees, each of which has height $\Oh(\log n)$.
	Specifically, the topmost level $\ttd(F)$ consists of the nodes corresponding to the internal vertices of $T$, and the previous levels are defined analogously by recursion (e.g., level $\ttd(F)-1$ consists of the union of the topmost levels of $\T_i$ over all $i \in [d]$, etc.).
	Thus, $\height(\T) \in \Oh (\ttd(F) \log(n))$.
	
	There are $n$ leaves of $\T$, one for each column of $A$.
	The initialization of leaves takes time at most $n \cdot (2\rho+1)$.
	Let $N_\ell$, $\ell \in [\ttd(F)]$, denote the number of internal nodes at level $\ell$.
	Because a full binary tree has at most as many internal nodes as it has leaves, we see that $\sum_{\ell=1}^{\ttd(F)} N_\ell \leq n$.
        Then, initializing level $\ttd(F)$ amounts to solving $N_{\ttd(F)}$ convolutions and their witnesses, each of which is computable in time $\bigO(|R'|^2) \leq \bigO\left(\left((2\rho + 1)\cdot (2\norm{A}_\infty \cdot \rho + 1)\right)^{2\height(F)}\right) \leq \bigO\left((2\norm{A}_\infty \cdot \rho + 1)^{2\height(F)+1}\right)$.
	Processing level $\ell \in [\ttd(F)-1]$ amounts to solving $N_\ell$ convolutions with sets $R'(\ell) = R(\ell) \times [0,\rho]$ where $R(\ell)$ is obtained from $R$ by dropping some coordinates, and thus~$|R'(\ell)| \leq |R'|$, and hence computing one convolution can be done in time $\bigO(|R'(\ell)|^2) \leq \bigO\left((2\|A\|_\infty \cdot \rho + 1)^{2\height(F)+1}\right)$.
	In total, initialization takes time at most $\bigO\left(2n (2\|A\|_\infty \cdot \rho + 1)^{2\height(F)+1}\right)$.
	
	Regarding the \textsc{Update} operation, there is one leaf of $\T$ corresponding to the coordinate $i$ being changed.
	Thus, in order to update the results, we need to recompute all convolutions corresponding to internal nodes along the path from the leaf to the root.
	Because $\height(\T) \leq \Oh(\ttd(F) \log n)$, the number of such nodes is $\Oh(\ttd(F)\log n)$, with each taking time at most $\bigO\left((2\|A\|_\infty \cdot \rho + 1)^{2\height(F)+1}\right)$.
	% The time required by $\sigma$-\textsc{Update} is $\sigma$ times the time required by one \textsc{Update} operation.

        As the output of~\textsc{Query} has a sparse representation described in the statement of the lemma, the running time for \textsc{Query} follows.
        \qed
\end{proof}

%% file: conclusion.tex
\subsection*{Open Problems}
Our results primarily point in two interesting research directions.
First, is it possible to eliminate the $\log n$ factor in the dual algorithm (Theorem~\ref{thm:dual-alg}), at least for some classes of~\ref{IP} with small $\td_D(A) + \|A\|_\infty$?
If not, how could the $n \log \|\veu-\vel\|_\infty$ lower bound be strengthened to match our results?
Second, the lower bound does not apply outside of the comparison oracle model; for which classes of objectives beyond linear can one design faster, possibly strongly polynomial algorithms?

%% file: convolution-tree-opt.tex
% TEX root = main.tex
\section{Optimality of the Convolution Tree}
\label{sec:convolution-tree-opt}

In this section we argue that our realization of the convolution tree (Lemma~\ref{lem:convol_tree}) is optimal with respect to~\(n\).
Formally, we prove the following lemma.
\begin{lemma}
  \label{lem:convolution-tree-opt}
  In any implementation of the convolution tree in the comparison model, the total running time of one operation~\textsc{Init}, \(n\)~operations~\textsc{Update} and~\(n\)~operations~\textsc{Query} requires time~\(\Omega(n \log n)\).
\end{lemma}
\begin{proof}
  We prove this lemma by demonstrating how to sort an array~\(\vealpha = (\alpha_1, \ldots, \alpha_n)\) of integers using the convolution tree with one \textsc{Init}, \(n\)~\textsc{Updates} and~\(n\)~\textsc{Queries}.

  We guide ourselves with the following~\ref{IP} instance.
  Let~\(x_1, \ldots, x_{n + 1}\) be $n+1$ integer variables.
  Let~\(C\) be a constant larger than~\(\sum_{i = 1}^n \alpha_i\).
  The~\ref{IP} instance shall be
  \begin{equation*}
  \begin{array}{lr@{}c@{}l}
    \text{minimize}    & C x_{n + 1} + \sum_{i = 1}^n \alpha_i x_i & \\
    \text{subject to } & \sum_{i = 1}^{n + 1} x_i                  & = n   \\
                       & x_1, \ldots, x_n                          & \le 1 \\
                       & x_{n + 1}                                 & \le n \\
                       & x_1, \ldots, x_{n + 1}                    & \ge 0
  \end{array}
  \end{equation*}
  Note that this instance has dual treedepth~\(1\) and~\(\norm{A}_\infty = 1\).
  We initialize a convolution tree~\(\T\) with~\(\vel = \vezero, \veu = (1, \ldots, 1, n)\) and~\(f(\vex) = C x_{n + 1} + \sum_{i = 1}^n \alpha_i x_i\).
  We will find an optimum solution by starting from a feasible solution~\(\vez^0 = (0, \ldots, 0, n)\).
  For~\(i \in [n]\) let~\(\vea^i \in \Z^{n + 1}\) be a vector which has only two non-zero entries, at index~\(i\) it has a~\(1\) and at index~\(n + 1\) it has~\(-1\).
  It is easy to see that the only minimum augmenting steps are~\(\vea^1, \ldots, \vea^n\).
  (Indeed, this is exactly the relevant subset of the Graver basis of $A$ mentioned in the previous section.)

  We start by looking for an augmenting step~\(\veh\) as in~\ref{l1-IP} with~\(\rho = 2\) by performing a \textsc{Query} operation on~\(\T\).
  Clearly each~\(\vea^i\) can be sparsely represented.
  At the beginning, the augmenting step~\(\veh\) that decreases the minimized objective function the most is~\(\veh^0 := \vea^{i_1}\) where~\(i_1\) is the index of the maximum element of~\(\vealpha\).
  We apply \textsc{Update} on~\(\T\) to re-center the instance at~\(\vez^1 = \vez^0 + \veh^0\).
  If we look for another augmenting step, we cannot use the same augmenting step~\(\veh^0\) as the constraint~\(x_{i_1} \le 1\) has become tight.
  Thus the next augmenting step~\(\veh^1\) will be~\(\vea^{i_2}\) where~\(i_2\) corresponds to the second largest entry~\(\alpha_{i_2}\) among~\(\vealpha\), and so on when we search for~\(\veh^2, \ldots, \veh^{n - 1}\).
  Eventually the sequence of augmenting steps~\(\vea^{i_1}, \ldots, \vea^{i_n}\) returned by $\T$ has the property that~\(i_j\) is the index of the~\(j\)\hy{}th largest entry of~\(\alpha\).
  Thus we have listed~\(\vealpha\) in non-increasing order.

  Our instance has~\(\norm{A}_\infty, \td_D(A), \rho \in \Oh(1)\), hence \textsc{Init} takes time~\(\Oh(n)\), \textsc{Update} time~\(\Oh(\log n)\), and \textsc{Query} time~\(\Oh(1)\).
  In total, we used one \textsc{Init}, \(n\)~\textsc{Updates}, and~\(n\)~\textsc{Queries}.
  If we could perform this combination of operations in time~\(o(n \log n)\), then we would get a contradiction with the known fact that sorting requires time~\(\Omega(n \log n)\) in the comparison model.
  Note that we only copy~\(\alpha_i\)'s, thus our lower bound indeed works in the comparison model.
  In particular, the convolutions that occur never involve the numbers~\(\alpha_i\).
  This concludes the proof.
  \qed
\end{proof}